\documentclass[usenatbib,fleqn]{mnras}
\usepackage{graphicx}
\usepackage{amssymb}
\usepackage{amsmath}
\usepackage{xcolor}
\newcommand{\Pa}{\mathcal{P}}

\newcommand{\mpchi}{\,h^{-1}{\rm {Mpc}}}

\newcommand{\Mh}{M_{\rm h}}
\newcommand{\Ac}{A_{\rm c}}
\newcommand{\Mc}{M_{\rm c}}
\newcommand{\sigc}{\sigma_{\rm c}}
\newcommand{\As}{A_{\rm s}}
\newcommand{\Ms}{M_{\rm s}}
\newcommand{\sigs}{\sigma_{\rm s}}

\newcommand{\rp}{r_{\rm p}}
\newcommand{\ngg}{n_{\rm g}}
\newcommand{\hinvMsun}{h^{-1}{\rm M}_\odot}
\newcommand{\hinvMpc}{h^{-1}{\rm Mpc}}

\newcommand{\Mr}{M_{\rm r}}
\newcommand{\Msun}{{\rm M}_\odot}
\newcommand{\wproj}{w_{\rm p}}
\newcommand{\alphac}{\alpha_{\rm c}}
\newcommand{\alphas}{\alpha_{\rm s}}

\title[Galaxy Assembly Bias in Velocity Space]{On the Constraints of Galaxy Assembly Bias in Velocity Space}

\author[K.S. McCarthy, Z. Zheng, H. Guo, W. Luo, and Y-T. Lin]{
\parbox{\textwidth}{
Kevin S. McCarthy$^{1}$\thanks{E-mail: kevmac@caltech.edu},
Zheng Zheng$^{1}$\thanks{E-mail: zhengzheng@astro.utah.edu},
Hong Guo$^{2}$,
Wentao Luo$^{3,4}$,
and
Yen-Ting Lin$^{5}$\\
}
\\
$^{1}$ Department of Physics and Astronomy, University of Utah, Salt Lake City, UT 84112, USA\\
$^{2}$ Key Laboratory for Research in Galaxies and Cosmology, Shanghai Astronomical Observatory, Shanghai 200030, China \\
$^{3}$ Kavli Institute for the Physics and Mathematics of the Universe (WPI), the University of Tokyo, Kashiwa, Chiba 277-8582, Japan\\
$^{4}$ CAS Key Laboratory for Research in Galaxies and Cosmology, Department of Astronomy, University of Science and Technology of China, Hefei, Anhui 230026, China\\
$^{5}$ Institute  of  Astronomy  and  Astrophysics,  Academia  Sinica, Taipei 10617, Taiwan
}

\begin{document}
\label{firstpage} \pagerange{\pageref{firstpage}--\pageref{lastpage}}
\maketitle
%%%%%%%%%%%%%%%%%%%%%%%%%%%%%%%%%%%%%%%%%%%%%%%%%%%%%%%%%%%%%%%%%%%%%%%%%%%%%%%%%%%%%%%%%%%%%%%%%
\begin{abstract}

If the formation of central galaxies in dark matter haloes traces the assembly history of their host haloes, in haloes of fixed mass, central galaxy clustering may show dependence on properties indicating their formation history. Such a galaxy assembly bias effect has been investigated previously, with samples of central galaxies constructed in haloes of similar mass and with mean halo mass verified by galaxy lensing measurements, and no significant evidence of assembly bias is found from the analysis of the projected two-point correlation functions of early- and late-forming central galaxies. In this work, we extend the the investigation of assembly  bias  effect  from  real  space to redshift (velocity) space, with an extended construction of early- and late-forming galaxies. We carry out halo occupation distribution modelling to constrain the  galaxy-halo connection to see whether there is any sign of the effect of assembly bias. We find largely consistent host halo mass for early- and late-forming central galaxies, corroborated by lensing measurements. The central velocity bias parameters, which are supposed to characterise the mutual relaxation between central galaxies and their host haloes, are inferred to overlap between early- and late-forming central galaxies. However, we find a large amplitude of velocity bias for early-forming central galaxies (e.g. with central galaxies moving at more than 50\% that of dark matter velocity dispersion inside host haloes), which may signal an assembly bias effect. A large sample with two-point correlation functions and other clustering measurements and improved modelling will help reach a conclusive result.

\end{abstract}

%%%%%%%%%%%%%%%%%%%%%%%%%%%%%%%%%%%%%%%%%%%%%%%%%
 
%%%%%%%%%%%%%%%%%%%%%%%%%%%%%%%%%%%%%%%%%%%%%%%%%%%%%%%%%%%%%%%%%%%%%%%%%%%%%%%%%%%%%%%%%%%%%%%%%%
\section{Introduction}
\label{sec:intro}

Galaxy clustering data from large-volume galaxy redshift surveys has been playing an important role in studying galaxy formation and evolution and in learning about cosmology. As galaxies form and evolve inside dark matter haloes and the halo population can be readily predicted by a given cosmology model (e.g. through cosmological $N$-body simulations), models based on the galaxy-halo connection have become the convenient and natural route to interpret galaxy clustering data. 

The commonly adopted descriptions of galaxy-halo relation (see \citealt{Wechsler18} for a review), such as the framework of halo occupation distribution (HOD; e.g. \citealt{Berlind03}) and conditional luminosity function (CLF; e.g. \citealt{Yang03}), have been successfully applied to galaxy surveys to infer the relation between galaxy properties and halo mass (e.g. \citealt{vandenBosch03,Zheng04,Zehavi05,Zehavi11,Guo14,XuHJ18}). With such modelling, the high precision clustering measurements at small to intermediate scales (i.e., sub-Mpc to tens of Mpc) have the potential to tighten cosmological constraints \citep[e.g.][]{Abazajian05,Zheng07,vandenBosch13,Reid14,Zhai19}. 

The implicit assumption in the above frameworks is that the statistical properties of galaxies inside haloes only depend on halo mass, not on halo environment or halo growth history, partly motivated by the excursion set theory \citep{Bond91}. However, it was later discovered, through $N$-body simulations \citep[e.g.][]{Sheth04,Gao05} and theoretical study of the statistics of the peaks of Gaussian random fluctuations \citep[e.g.][]{Dalal08}, that the clustering of haloes depends on not only halo mass but also halo assembly history, which is termed halo assembly bias and has been under active investigations \citep[e.g.][]{Paranjape18,Xu18,Han19}. If in haloes of fixed mass galaxy properties and halo assembly history are closely connected, the inherited assembly bias from the host haloes would be manifested as effects on galaxy clustering. While such galaxy assembly bias, which can be conceptually defined as dependence of galaxy clustering on galaxy properties at fixed halo mass, could enable us to learn more about galaxy formation and evolution, it may introduce possible systematics in cosmological constraints \citep[e.g.][]{Zentner14} or limit the range of clustering data to be used \citep[e.g.][]{McCarthy2019} if not properly accounted for.

An empirical detection of galaxy assembly bias effect will advance our understanding on galaxy formation and also guide the model development for cosmological constraints with galaxy clustering.  No definite conclusion has been reached in the search of galaxy assembly bias effect in galaxy survey data. Analyses of projected galaxy two-point correlation functions (2PCFs) in the Sloan Digital Sky Survey (SDSS) show no significant evidence of galaxy assembly bias \citep[e.g.][]{Lin16,Vakili19,Zentner19,Salcedo20}. The reported strong assembly bias effect in the projected 2PCFs of massive clusters \citep[][]{Miyatake16} appears to be an artefact caused by an insufficient control of projection effects \citep{Zu17}. Based on gravitational lensing measurements, \citet{Zu21} find a $\sim 10$ per cent difference in the large scale bias between low stellar mass, low concentration and high stellar mass, high concentration galaxy clusters of the same average mass, and it remains to be investigated how much this difference in bias can be explained by the difference in the cluster mass distribution. While \citet{Yuan21} claim the detection of galaxy assembly bias in redshift-space clustering of BOSS CMASS galaxies, it is in contradiction to the result in \citet{Guo15a}. 

One major difficulty in searching for assembly bias effect in observational data lies in the halo mass determination, so that assembly bias effect can be well separated from the effect of clustering from the halo-mass dependence. In such regard, the assembly bias analysis in \citeauthor{Lin16} (\citeyear{Lin16}; hereafter L16) is distinct. They construct samples of central galaxies in SDSS in haloes of similar mass, and the halo mass is verified by galaxy lensing measurements. Their analysis with the projected 2PCFs of central galaxy samples with different galaxy growth history and star formation properties shows no significant evidence for assembly bias, which suggests that the correlation, if any, between the chosen galaxy assembly indicator and halo formation history may not be tight \citep{Xu20}.

The L16 analysis focuses on projected 2PCFs, effectively galaxy clustering in real space. In this work, we extend the clustering analysis in L16 to redshift space with an extended sample construction. Modelling of luminosity-threshold samples of galaxies in BOSS and SDSS has revealed the existence of velocity bias \citep{Guo15a,Guo15b}, which is the difference between the motion of galaxies and dark matter particles inside haloes. In particular, the central galaxies do not appear to be at rest with respect to the host haloes (e.g., in the halo center-of-mass frame), moving with a mean velocity about 20--30 per cent of the velocity dispersion of dark matter inside haloes, which reflects the mutual relaxation of galaxies and haloes \citep[][]{Ye17}. Motivated by the expectation that central galaxies of different assembly/formation histories may have different degrees of relaxation with respect to host haloes, we perform investigations of the redshift-space clustering with the extended L16 samples to see whether such a difference exists or whether there is any hint for assembly bias in velocity space.

This paper is organised as follows. In Section~\ref{sec:data}, we present the details of the construction of the various galaxy samples, extending those in L16. In Section~\ref{sec:method}, we describe the clustering measurements and the modelling method. The results are presented in Section~\ref{sec:Results}, followed by a discussion in Section~\ref{sec:discussion}. Finally in Section~\ref{sec:conc}, we summarise the work and give the conclusion.

%%%%%%%%%%%%%%%%%%%%%%%%%%%%%%%%%%%%%%%%%%%%%%%%%%%%%%%%%%%%%%%%%%%%%%%%%%%%%%%%%%%%%
\section{Data}
\label{sec:data}

For studying galaxy assembly bias, we would like to have samples of central galaxies in haloes of fixed mass to investigate the dependence of clustering on galaxy properties. L16 devise a procedure to construct early-forming and late-forming galaxy samples to have similar mean host halo mass of central galaxies, which is verified with galaxy lensing measurements. For simplicity, in this work the early-forming (late-forming) galaxies are referred to as the early (late) galaxies.

This work will entail an extended analysis of the early/late galaxy samples in L16. The galaxies in these samples are supposed to be central galaxies, selected base on the group catalogue of \citeauthor{Yang07} (\citeyear{Yang07}; Y07) constructed from SDSS (\citealt{York00}) Data Release 7 (DR7; \citealt{Abazajian09}), specifically from DR7 modelC\footnote{\url{https://gax.sjtu.edu.cn/data/Group.html}}. We use `LIN' to label the original catalogue used in L16. 

L16 applied a satellite decontamination algorithm to remove the incorrectly labelled central galaxies from Y07. As will be shown in section~\ref{SFHres}, this algorithm also removes a fraction of central galaxies, when they become close in the direction transverse to the line of sight. It causes a paucity of signal at small scales for the projected two-point correlation function. While this has no effect on the L16 results based on large-scale projected clustering, it would affect our analysis of redshift-space clustering. Therefore we also have samples of galaxies selected with the same criteria as in L16 but without the satellite decontamination algorithm. These samples  are labelled as `PCAT', for parent catalogues. The satellite component will be accounted for in our model.

Finally, to increase the signal-to-noise ratio of the clustering measurements and hence improve the model constraints, we also construct galaxy catalogues by loosening the selection criteria, which extends the halo mass range to include more galaxies. We use `EXT' to denote these extended samples. 

In what follows, we describe the selection of the galaxy samples and the division into the early and late galaxy samples according to star formation history (SFH) and specific star formation rate (sSFR). The galaxy counts and number density of each galaxy sample are summarised in Table~\ref{table:gals}.

\begin{table*}
    \caption{
    Early/late galaxy samples constructed based on SFH and sSFR. Listed are the total number ($N$) of galaxies and the comoving number density ($n_{\rm g}$; in units of $10^{-3}h^3{\rm Mpc^{-3}}$) for each sample. Also listed is the mean halo mass for central galaxies ($\langle M_{\rm h}\rangle$; in units of $\hinvMsun$), inferred in this paper from modelling the projected correlation function [$w_{\rm p}$] only and from modelling both the projected and redshift-space two-point correlation functions [$w_{\rm p}$, $\xi_0$, $\xi_2$] (see the text for details).
    }
    \centering
    \begin{tabular}{lcccccccc}
     \hline                    %inserts double horizontal lines
        & \multicolumn{4}{c}{SFH-based} &  \multicolumn{4}{c}{sSFR-based}\\
     \hline
Sample & $N$ & $n_{\rm g}$ & $\log\langle M_{\rm h}\rangle |_{[w_{\rm p}]}$ &  $\log\langle M_{\rm h}\rangle |_{[w_{\rm p},\xi_0,\xi_2]}$ & $N$ & $n_{\rm g}$ & $\log\langle M_{\rm h}\rangle |_{[w_{\rm p}]}$ & $\log\langle M_{\rm h}\rangle |_{[w_{\rm p},\xi_0,\xi_2]}$\\
\hline % inserts single horizontal line  also add wp fitting results
LIN--early  & 18199 & 0.53 & $11.64_{-0.34}^{+0.28}$ & $11.82_{-0.14}^{+0.11}$ & 25837 & 0.60 & $12.57_{-0.18}^{+0.12}$ & $12.43_{-0.05}^{+0.07}$\\ % inserting body of the table
LIN--late   & 26070 & 0.24 & $12.23_{-0.34}^{+0.17}$ & $12.22_{-0.08}^{+0.07}$ & 29658 & 0.21 & $12.22_{-0.38}^{+0.16}$ & $12.24_{-0.09}^{+0.08}$\\
PCAT--early & 19711 & 0.64 & $11.78_{-0.41}^{+0.27}$ & $12.04_{-0.20}^{+0.14}$ & 29418 & 0.74 & $12.65_{-0.21}^{+0.13}$ & $12.62_{-0.08}^{+0.07}$\\ % inserting body of the table
PCAT--late  & 30549 & 0.30 & $12.51_{-0.32}^{+0.16}$ & $12.45_{-0.09}^{+0.08}$ & 33420 & 0.23 & $12.36_{-0.25}^{+0.14}$ & $12.44_{-0.12}^{+0.12}$\\
EXT--early  & 59871 & 1.40 & $12.20_{-0.29}^{+0.13}$ & $12.22_{-0.06}^{+0.04}$ & 52014 & 1.56 & $12.21_{-0.10}^{+0.08}$ & $12.20_{-0.05}^{+0.05}$\\ % inserting body of the table
EXT--late   & 37684 & 0.50 & $12.48_{-0.23}^{+0.13}$ & $12.32_{-0.06}^{+0.06}$ & 37746 & 0.22 & $12.22_{-0.12}^{+0.09}$ & $12.18_{-0.05}^{+0.11}$\\
\hline %inserts single line
    \end{tabular}
    \label{table:gals}
\end{table*}

\subsection{SFH-based Construction}\label{sec:SFHsel}

We first describe how to form early/late galaxy samples with the SFH information of galaxies. The construction makes use of galaxy stellar mass, galaxy colour, and galaxy SFH.

Stellar mass is estimated based on absolute $r$-band magnitude ($\Mr$) and colour ($g-r$) with the empirical formula of \citet{Bell03}, 
\begin{equation}
\log\left(\frac{M_*}{h^{-2}\Msun}\right)=-0.406+1.097(g-r)-0.4(\Mr-5\log h-4.64), \label{eq:bell}
\end{equation}
where $\Mr$ is the Petrosian magnitude $K$-corrected to $z=0$ from the NYU value-added galaxy catalogue \citep{Blanton05}. Galaxies are divided into red and blue types following a dividing line defined as 
\begin{equation}
(g-r)_{\rm div}=0.67+0.03\log\left[M_*/(10^{10}h^{-2}\Msun)\right].
\label{eq:colorcut}
\end{equation}

Galaxy SFH comes from the versatile spectral analysis (VESPA;\citealt{Tojeiro07,Tojeiro09})\footnote{\url{http://www-wfau.roe.ac.uk/vespa/}}, which adopts the stellar population synthesis model of \citet{Bruzual03}, with the stellar initial mass function of \citet{Chabrier03} and the Padova stellar evolution tracks \citep{Alongi93,Bressan93,Fagotto94a,Fagotto94b,Girardi96}. The data with the highest temporal resolution are used, where the SFR is stored in 16 logarithmically-spaced bins with upper bin boundary ranging from 0.02 to 14 Gyr in lookback time (fig.1 of \citealt{Tojeiro09}). We divide the stellar mass formed in the earliest bin of lookback time (from 9 to 14 Gyr, or $z \gtrsim 1.35$) by the total stellar mass. If this fraction is  $\ge 0.5$ ($< 0.5$), the galaxy is classified as early-formed (late-formed). VESPA employs two dust extinction models, and we only keep galaxies with consistent classifications between the two models.

When constructing early and late galaxy samples with similar mean halo mass, L16 find that there is a systematic effect in the halo mass estimates in the Y07 group catalogue, as the colour dependent relation between halo mass and stellar mass \citep{More11} is not taken into account. After accounting for such a dependence, L16 use different stellar mass ranges for red and blue galaxies to select halos of similar mass range. L16 further find that the SFH of galaxies (early-formed/late-formed) also need to be considered. L16 choose four ranges of stellar mass, here denoted as $\mathcal{M}_i$ ($i=$1, 2, 3, and 4), corresponding to $\log[M_*/(h^{-2}\Msun)]=[9.9, 10.2]$, [10.2, 10.45], [10.35, 10.6], and [10.52, 11.1], respectively. 
With the sets of red/blue, $\mathcal{M}_i$, and early-formed/late-formed galaxies, the construction of samples of early and late galaxies in L16 can be expressed as
\begin{equation}
{\rm Early}= \left[({\rm Red} \cap \mathcal{M}_1) \cup ({\rm Blue} \cap \mathcal{M}_2)\right] \cap \text{Early-formed}
\end{equation}
and
\begin{equation}
{\rm Late}= \left[({\rm Red} \cap \mathcal{M}_3) \cup ({\rm Blue} \cap \mathcal{M}_4)\right] \cap \text{\rm Late-formed},
\end{equation}
where `$\cup$' and `$\cap$' represent the union and intersection of sets, respectively.

With the above construction procedure, the LIN catalogue (with satellite decontamination using a friends-of-friends algorithm) has 18199/26070 galaxies in the early/late sample. Consistent mean halo masses are found from galaxy lensing measurements in L16, $M_{\rm 200c}=9.5^{+2.5}_{-2.0}\times 10^{11} h^{-1}\Msun$  and $8.4^{+2.2}_{-1.8} \times 10^{11} h^{-1}\Msun$ for the early and late sample, respectively, where $M_{\rm 200c}$ is the halo mass within a radius within which the mean density is 200 times the critical density of the universe.

Our PCAT catalogue, which has the same selection criteria as in L16 but without the satellite decontamination, has 19711/30549 galaxies in the early/late sample. In order to improve the signal-to-noise ratio of clustering measurements, we increase the number of galaxies to form the EXT catalogue by slightly extending the stellar mass ranges, with $\mathcal{M}_i$ ($i=$1, 2, 3, and 4) changing to $\log[M_*/(h^{-2}\Msun)]=[9.9, 10.4]$, [10.2, 10.75], [10.2, 10.6], and [10.5, 11.1], respectively. This results in 59871/37684 galaxies in the early/late sample.

\subsection{sSFR-based Construction}
\label{sec:sSFRsel}

In L16, the second way to constructing early and late galaxy samples is to use the sSFR. Galaxies with higher sSFR are still actively forming stars and therefore have more of their stellar mass formed recently, while galaxies with lower sSFR have formed more of their stellar mass at earlier time.

The values of sSFR in Y07 are obtained from the The Max Planck Institute for Astrophysics and Johns Hopkins University (MPA-JHU) value-added catalogues, where a modified version of the \cite{Brinchmann04} technique has been employed. L16 find that different combinations of Y07 halo mass ($M_{\rm 200c}$) range and sSFR values can result in consistent mean halo mass, as verified by galaxy lensing measurements. The sample of early galaxies are selected to have low sSFR, $\log({\rm sSFR}/{\rm yr}^{-1})< -11.8$, in a Y07 halo mass range of $\log(M_{\rm 200c}/\Msun)$=12.0--12.5. For the sample of late galaxies, we have $\log({\rm sSFR}/{\rm yr}^{-1})> -11.8$ and $\log(M_{\rm 200c}/\Msun)$=12.75--13.1. With satellite decontamination, the LIN catalogue ends up with 25837/29658 galaxies in the early/late sample. The mean halo mass estimates from galaxy lensing are $1.39^{+0.24}_{-0.21}\times 10^{12}h^{-1}\Msun$ and $1.26^{+0.24}_{-0.20}\times 10^{12}h^{-1}\Msun$ for the two samples.

Our PCAT catalogue without the satellite decontamination has 29658/33420 galaxies in the early/late sample. Similar to the SFH-based construction, we construct an EXT catalogue to increase the number of galaxies for improving the signal-to-noise ratio of the clustering measurements. With more galaxies, we are also able to shift the sSFR cut from $\log({\rm sSFR}/{\rm yr}^{-1})=-11.8$ to $-11.0$ to better separate the quenched and active star-forming galaxies. For the early EXT sample, we have $\log({\rm sSFR}/{\rm yr}^{-1})< -11.0$ in a Y07 halo mass range of $\log(M_{\rm 200c}/\Msun)$=11.8--12.3, and for the late EXT sample, $\log({\rm sSFR}/{\rm yr}^{-1})> -11.0$ with $\log(M_{\rm 200c}/\Msun)$=12.6--13.2. This results in 52014/37746 galaxies in the early/late sample in the EXT catalogue.

\section{Methodology}
\label{sec:method}

\subsection{Galaxy Clustering Measurements}
\label{sec:clusmeas}

The primary measurements considered in this work are the projected two-point correlation functions (2PCFs) $\wproj$ and the multipoles $\xi_l$ of the redshift-space 2PCFs. To obtain $\wproj$, we measure the three-dimensional (3D) 2PCF $\xi(\rp,r_\pi)$ in terms of the line-of-sight and transverse separations, $\rp$ and $r_\pi$, of galaxy pairs. To compute $\xi_l$, we measure the 3D 2PCF $\xi(s,\mu)$ in terms of the pair separation $s$ and the cosine of the angle between the line-of-sight direction and the pair separation vector. In detail, for two galaxies located at $\mathbf{v}_1$ and $\mathbf{v}_2$ in redshift space, we define the line-of-sight vector $\mathbf{l}\equiv(\mathbf{v}_1+\mathbf{v}_2)/2$ and separation vector $\mathbf{s}=\mathbf{v}_2-\mathbf{v}_1$ \citep[e.g.][]{Zehavi11}. We then have
\begin{equation}
s=|\mathbf{s}|,\,\,\,\,\, \mu\equiv \cos\theta= \mathbf{s}\cdot\mathbf{l}/|\mathbf{s}||\mathbf{l}|.
\label{eq:sueq}
\end{equation}
and 
\begin{equation}
r_\pi \equiv |\mathbf{s}\cdot\mathbf{l}|/|\mathbf{l}|,  \,\,\,\,\, \rp \equiv \sqrt{s^2-r_\pi^2}.
\label{eq:rppieq}
\end{equation}

We use the Landy-Szalay estimator \citep{Landy93} to measure $\xi(\rp,r_\pi)$ and $\xi(s,\mu)$,
%%%%%%%%%%%%%%%%%%%%%%%
\begin{equation}
\xi=\frac{\rm DD-2DR+RR}{\rm RR},
\label{eq:LSest}
\end{equation}
%%%%%%%%%%%%%%%%%%%%%%%
where DD, DR, and RR are the normalised numbers of data–data, data–random, and random–random galaxy pairs in the corresponding ($\rp$, $r_\pi$) or ($s$, $\mu$) bins. The random catalogue for each sample has the same footprint and redshift distribution as the sample, with about 5 million points. We adopt uniform $\rp$ bins in logarithmic space with a bin width $\Delta\log \rp = 0.2$, with $\log[\rp/(\mpchi)]$ ranging from $-1$ to $1.4$. The $r_\pi$ bins are uniform in linear space from 0 to 40$\mpchi$ with a bin width $\Delta r_\pi =2\mpchi$. The setup of $s$ bins are the same as $\rp$, and 20 linear bins are used for $\mu$ from 0 to 1 (i.e. with a bin width $\Delta\mu=0.05$).

The projected 2PCF $\wproj$ is calculated as
\begin{equation}
\wproj(\rp)=2\sum_i\xi(\rp,r_{\pi,i})\Delta r_\pi.
\end{equation}
The redshift-space 2PCF multipoles $\xi_l$ is calculated as
\begin{equation}
\xi_l(s)=(2l+1)\sum_i\xi(s,\mu_i) \Pa_l(\mu_i)\Delta\mu,
\label{eq:expand}
\end{equation}
where $\Pa_l$ is the $l$-th Legendre polynomial. We limit our analysis and modelling to the monopole $\xi_0$ and the quadrupole $\xi_2$, as the measurements of the hexadecapole $\xi_4$ are usually noisy and contains little information about velocity bias we are after. 

In comparison to L16, we extend the sample construction and make 2PCF measurements in redshift space. In addition, we apply a correction to the fibre collision effect. In SDSS, because of the finite size of the fibre plugs, two fibres could not be placed at a separation smaller than $55\arcsec$, corresponding to a projected separation of $\sim 0.1\hinvMpc$ at the median redshift $z\sim 0.12$. The effect extends to larger scales in redshift space with the Finger-of-God (FoG) effect.  While fibre collision has little influence on the large-scale $\wproj$ analysis in L16, it can become important for small-scale and redshift-space clustering analyses presented in this work. We employ the technique developed by \citet{Guo12} to correct the fibre collision effect. In short, it makes use of galaxy pairs in tile overlap regions to recover the counts of collided pairs. It has been successfully applied in previous work \citep[e.g.][]{Guo15a,Guo15b}. 

Finally, we also perform galaxy-galaxy lensing measurements for the galaxy samples. Following \citet{Wentao17}, from the galaxy shear measurements we obtain the excess surface density (ESD) profile for each galaxy sample,
\begin{equation}
\Delta\Sigma(R)=\bar{\Sigma}(<R)-\Sigma(R),
\label{eq:esd}
\end{equation}
where $\bar{\Sigma}(<R)$ is the mean mass surface density within a projected radius $R$ around galaxies in the sample and $\Sigma(R)$ is the mass surface density at radius $R$. We note that these are all new ESD measurements, and even for the LIN samples, they are independent of those in L16 (with slightly different cosmological models adopted). The ESD measurements are not included in our modelling. Instead, they are used to provide a cross-check of the modelling results. 

\subsection{Modelling the Clustering Measurements}
\label{sec:hod}

The projected and redshift-space galaxy 2PCFs are modelled within the HOD framework. Here we introduce the HOD parameterisation and the 2PCF calculations.

While the construction of galaxy samples focuses on selecting central galaxies, they include a small fraction of satellites, in particular for the samples without satellite decontamination. So we parameterise the mean occupation function of galaxies as the sum of central and satellite components. 

For each of the late/early samples, the central galaxies are supposed to reside in haloes within a small range of mass. Different from the step-like function for a luminosity-bin sample \citep[e.g.][]{Zheng05}, the mean occupation function of central galaxies is similar to that for a luminosity-bin sample or follows closely to the form of the conditional luminosity function of central galaxies \citep[e.g.][]{Yang08}. We model the central component of the mean occupation function for each sample as a Gaussian bump in terms of the logarithmic halo mass,
\begin{equation}
\langle N_{\rm cen}(\Mh) \rangle = \Ac\exp\left[-\frac{(\log \Mh-\log \Mc)^2}{2\sigc^2}\right],
\label{eq:Ncen}
\end{equation}
where the three parameters $\Ac$, $\log\Mc$, and $\sigc$ represent the amplitude, centre, and width of the Gaussian profile.

For satellites, with the SFH-based construction they have a stellar mass range similar to that of central galaxies. It implies that they live in haloes more massive than the hosts of central galaxies. With the sSFR-based construction, a halo mass cut is imposed based on the Y07 halo mass estimate, so the satellites may reside in haloes of the same mass range as central galaxies. It is also possible that they are in more massive haloes, given the inaccuracy in the Y07 halo mass estimate. As our purpose of the study is on central galaxies, the satellite parameters are treated as nuisance parameters. We perform tests to model the satellite mean occupation function with either a power-low form or a Gaussian form, and we find consistent constraints on the HOD of central galaxies. We present the results by adopting a Gaussian form of the satellite mean occupation function, similar to that of the central galaxies, with the three parameters being $\As$, $\log\Ms$, and $\sigs$,
\begin{equation}
\langle N_{\rm sat}(\Mh) \rangle = \As\exp\left[-\frac{(\log \Mh-\log \Ms)^2}{2\sigs^2}\right].
\label{eq:Nsat}
\end{equation}

The mean occupation function for all the galaxies is 
\begin{equation}
\langle N(\Mh) \rangle = \langle N_{\rm cen}(\Mh) \rangle + \langle N_{\rm sat}(\Mh) \rangle.
\label{eq:Nall}
\end{equation}
We jointly model the 2PCFs of early and late galaxies. As we parameterise the mean occupation functions of early and late central galaxies separately, there is a possibility that the sum of the mean occupation numbers of early and late central galaxies exceeds unity, which becomes nonphysical. In such cases, we renormalise the amplitudes of the two samples, $A_{\rm c, early}\leftarrow A_{\rm c, early}/(A_{\rm c, early}+A_{\rm c, late})$ and $A_{\rm c, late}\leftarrow A_{\rm c, late}/(A_{\rm c, early}+A_{\rm c, late})$.

To model the redshift-space clustering, we also have two velocity bias parameters ($\alphac$ and $\alphas$) to describe the differences between the motions of galaxies and dark matter \citep[e.g.][]{Guo15a}. In the rest frame (centre-of-mass frame) of a halo, the central galaxy is not necessarily at rest, and we assume its line-of-sight velocity to follow a Laplacian distribution \citep{Guo15c} with velocity dispersion $\sigma_{\rm v, c}=\alphac \sigma_{\rm v, h}$. Here $\sigma_{\rm v, h}$ is the one-dimensional (1D) velocity dispersion of dark matter particles inside the halo, and the central velocity bias parameter $\alphac$ characterises the status of mutual relaxation between central galaxies and their host haloes \citep[e.g.][]{Ye17}. For satellites, we parameterise their line-of-sight velocity in the rest frame of a halo to follow a Gaussian distribution, characterised by the velocity dispersion $\sigma_{\rm v, s}=\alphas \sigma_{\rm v, h}$, with $\alphas$ the satellite velocity bias parameter. In the case that galaxies follow the motion of dark matter particles, we have $\alphac=0$ and $\alphas=1$. Our main purpose of this study is to see whether there is difference between the motions of early and late central galaxies. That is, we focus on the constraints on $\alphac$, and the satellite velocity bias $\alphas$ is treated as a nuisance parameter.

With the HOD parameters, we compute the model 2PCFs following the simulation-based method developed in \cite{Zheng16}. All the necessary information in fine halo mass bins relevant to 2PCF calculations is tabulated based on haloes identified in an $N$-body simulation. With the tables, the galaxy 2PCFs are obtained through summation over halo mass bins with galaxy occupation distribution accounted for. The method allows for efficient and accurate clustering predictions without the need to create mock galaxy catalogues. In this work, the tables are computed with {\sc rockstar} haloes \citep{Behroozi13a} from the {\sc MultiDark MDPL2} simulation \citep{Klypin16}. The simulation assumes a spatially flat cosmology with $\Omega_{\rm m} = 0.307$, $\Omega_{\rm b} = 0.048$, $h = 0.678$, $n_{\rm s} = 0.96$, and $\sigma_8 = 0.823$. The simulation has a box size of $1 h^{-1}$Gpc (comoving) on a side with $3840^3$ particles, and the corresponding mass resolution is $1.51 \times 10^9 \hinvMsun$.

Given a set of HOD parameters, the predicted 2PCFs are equivalent to the average measurements from different realizations of mock catalogues. In each mock catalogue, central galaxies are placed at the centres of host haloes, and random dark matter particles inside haloes are selected to represent satellites. The possibility of a halo of mass $\Mh$ to host a central galaxy follows $\langle N_{\rm cen}(\Mh)\rangle$ (equation~\ref{eq:Ncen}), and the number of satellites follows a Poisson distribution with mean $\langle N_{\rm sat}(\Mh)\rangle$ (equation~\ref{eq:Nsat}). The velocities of central and satellites are modified according to the velocity bias parameters $\alphac$ and $\alphas$.

We employ the Markov Chain Monte Carlo (MCMC) to explore the HOD parameter space from modelling the 2PCFs and galaxy number densities. The likelihood function is based on the value of $\chi^2$. For each sample (early or late galaxies), the $\chi^2$ is constructed as
\begin{equation}
    \chi^2= \bmath{(\xi-\xi^*)}^{\rm T} \bmath{\mathsf{C^{-1}}}\bmath{(\xi-\xi^*)}
       +\frac{(\ngg-\ngg^*)^2}{\sigma_{\ngg}^2},
\label{eq:chi2}
\end{equation}
where $\bmath{\xi}$ is the 2PCF data vector with covariance matrix $\bmath{\mathsf{C}}$, $\ngg$ is the galaxy number density with an uncertainty $\sigma_{\ngg}$ (assumed to be $10\%$ of the measured value shown in Table~\ref{table:gals}). Quantities with (without) a superscript `$*$' represents the measurements (model predictions). We perform joint modelling, with the total $\chi^2$ the sum of those pertaining to the early and the late galaxy sample.

The covariance matrix of each sample is estimated with the jackknife sampling technique, with $101$ and $400$ jackknife samples for the LIN/PCAT and EXT sample, respectively. To account for the mean bias in the precision matrix (inverse of the covariance matrix), we employ a correction to the precision matrix following \citet{Hartlap07}. That is, the precision matrix is multiplied by $(n-p-2)/(n-1)$, where $n$ is the number of jackknife samples and $p$ the number of 2PCF data points. The correlations between 2PCFs of early and late galaxy samples are neglected, as they cannot be reliably estimated with the limited sample size.

Finally, as we measure galaxy-galaxy lensing signals for each galaxy sample, we also compare the lensing measurements to the predictions from the best-fit model as a consistency check. For such a comparison, we construct mock galaxy catalogues according to the best-fit model. With dark matter particles in the MDPL2 simulation box, we directly measure the ESD profile (equation~\ref{eq:esd}) around each mock galaxy and take the average. To speed up the calculation without losing precision, we down-sample the dark matter particles by randomly selecting 1\% of them. For future  investigations it would be beneficial to include the lensing measurements in addition to the galaxy clustering measurements to constrain model parameters. 

%%%%%%%%%%%%%%%%%%%%%%%%%%%%%%%%%%%%%%%%%%%%%
\begin{figure*}
\includegraphics[width=\textwidth]{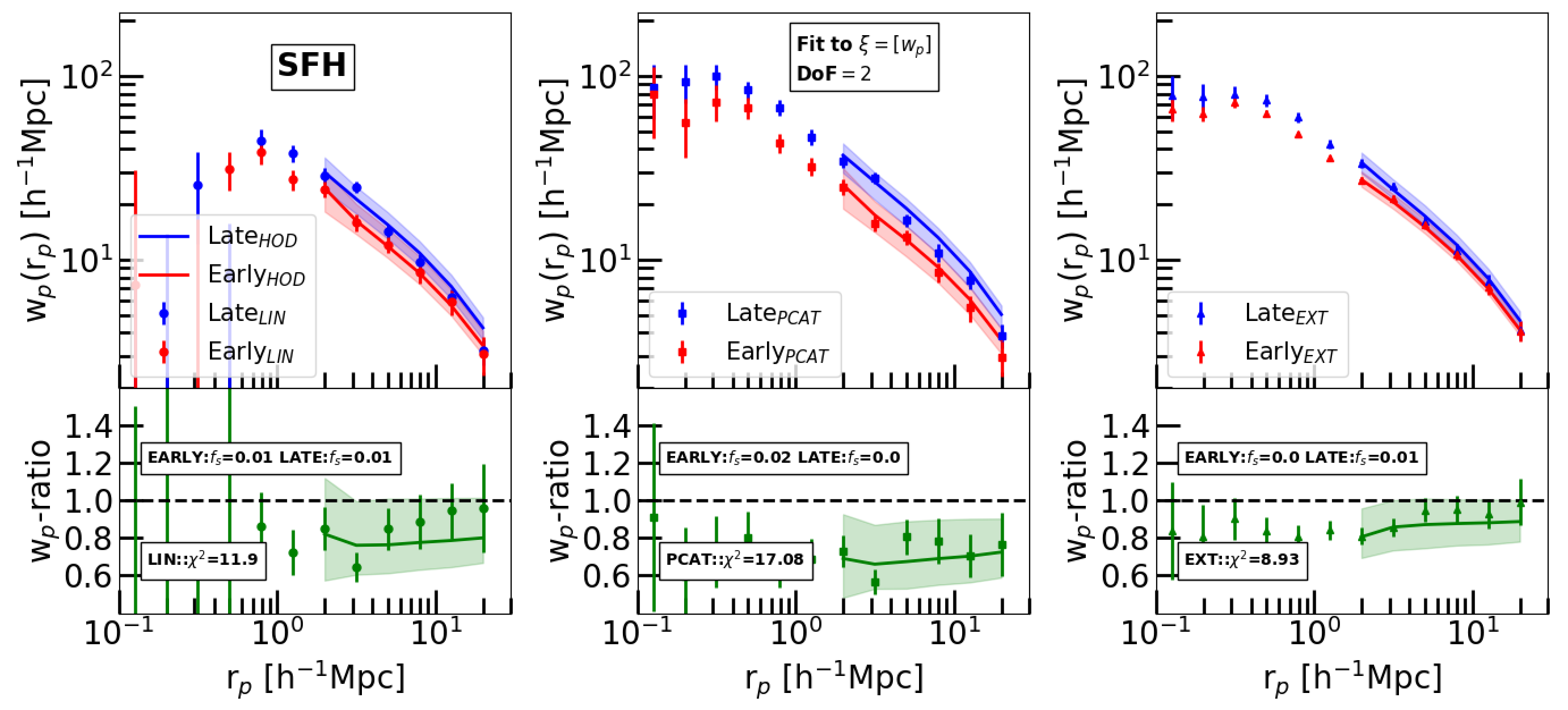}
\caption{
Measurements and modelling results of the projected 2PCFs $\xi=[\wproj]$ for the LIN, PCAT, and EXT samples (from left to right), with an SFH-based division into early and late galaxies. Measurements are shown as points in upper panels with jackknife error bars, and the predictions from the best-fitting HOD model (fitting to observations above $2\hinvMpc$) are shown as solid curves with the shaded region indicating the $1\sigma$ range. Ratios of early-to-late galaxy $\wproj$ measurements and best model fits are in the lower panels, where the $\chi^2$ values of the best fits (with 2 degrees of freedom) and the resultant satellite fractions $f_s$ are also displayed. 
}
\label{fig:SFHwp}
\end{figure*}
%%%%%%%%%%%%%%%%%%%%%%%%%%%%%%%%%%%%%%%%%%%%%%%%
%%%%%%%%%%%%%%%%%%%%%%%%%%%%%%%%%%%%%%%%%%%%%
\begin{figure*}
\includegraphics[width=\textwidth]{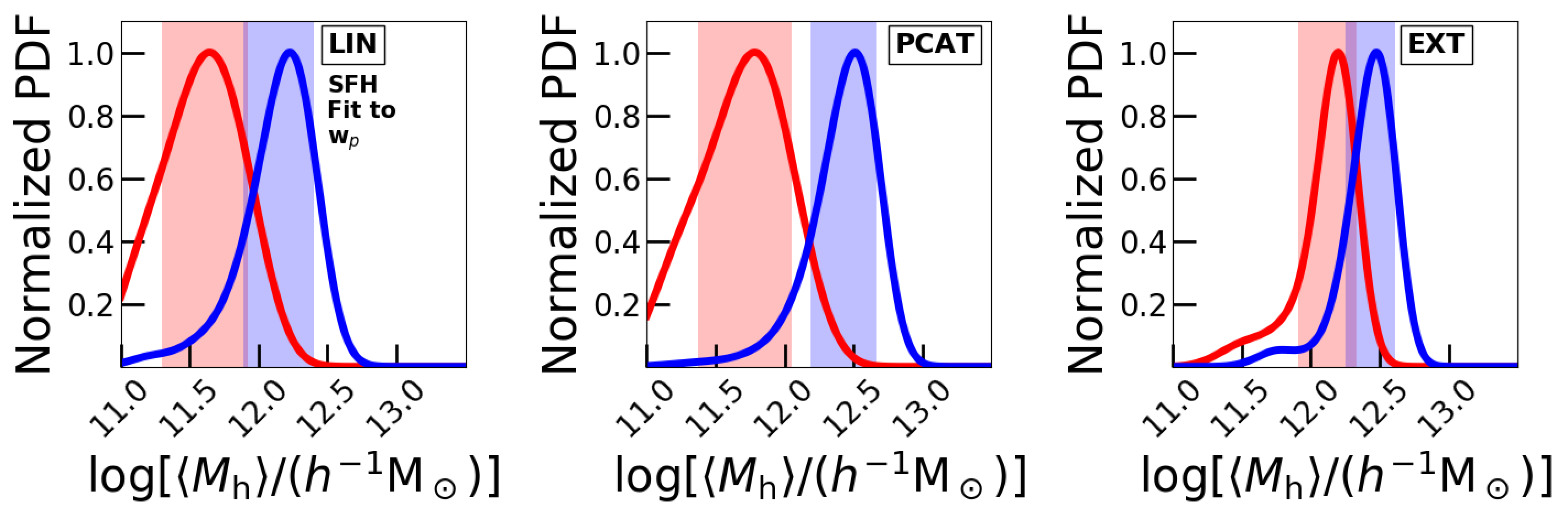}
\caption{Constraints on mean mass of central galaxy hosting haloes from modelling only the projected 2PCFs $\xi=[\wproj]$ for the LIN, PCAT, and EXT samples (from left to right), with an SFH-based division into early and late galaxies. For each sample, the curve show the PDF of the constraints, with the shaded region marking the central 68.3\% distribution. To characterise the difference between the distributions for early and late galaxies, we compute the significance values of the split between the peaks, which are 1.34$\sigma$ (LIN), 1.74$\sigma$ (PCAT), and 1.06$\sigma$ (EXT). The mean mass constraints are listed in Table~\ref{table:gals}.
}
\label{fig:SFHmassCONwp}
\end{figure*}
%%%%%%%%%%%%%%%%%%%%%%%%%%%%%%%%%%%%%%%%%%%%%%%%

%%%%%%%%%%%%%%%%%%%%%%%%%%%%%%%%%%%%%%%%%%%%%
\begin{figure*}
\includegraphics[width=\textwidth]{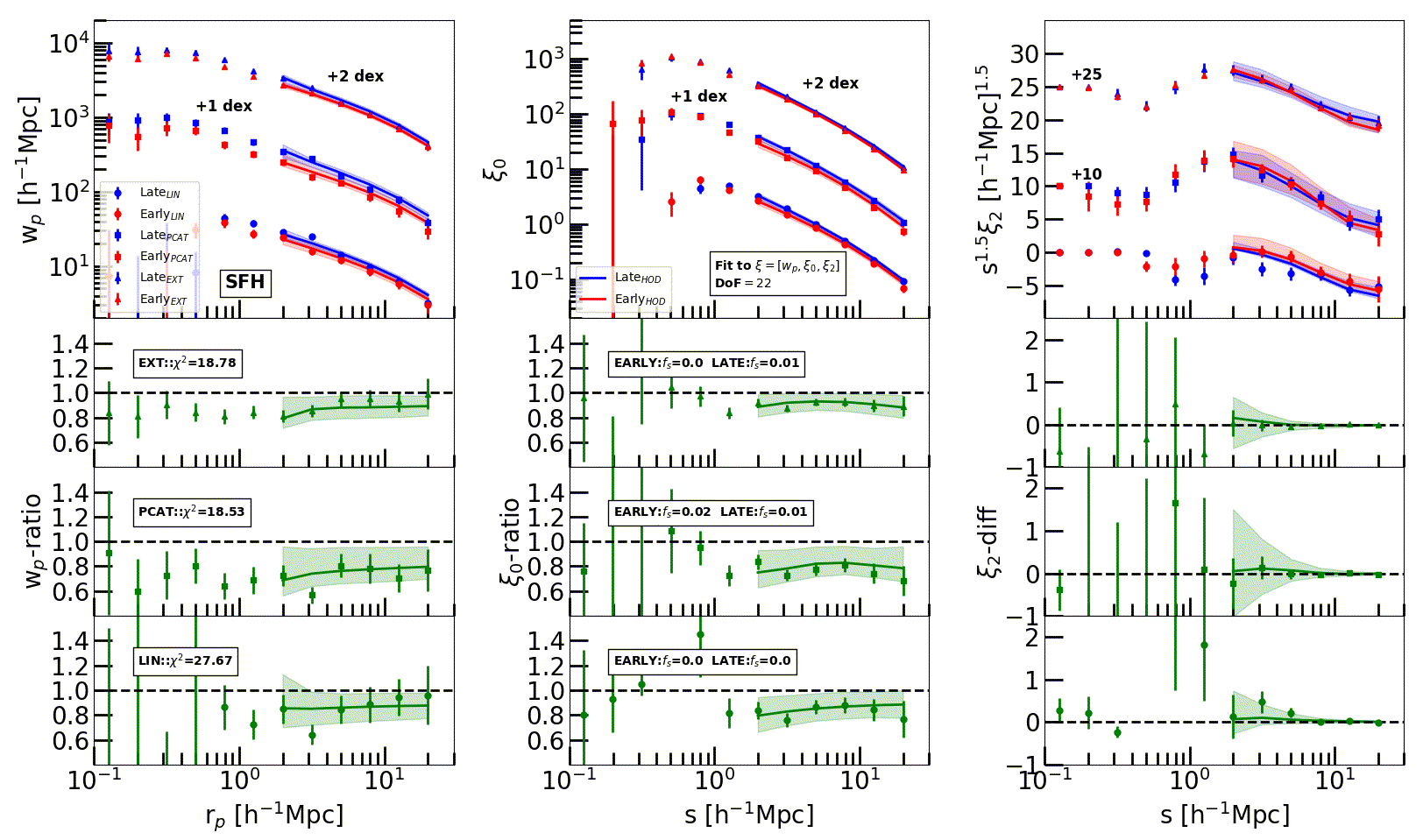}
\caption{
Measurements and modelling results of the projected 2PCFs and redshift-space multipoles, $\xi=[\wproj, \xi_0, \xi_2]$ (from left to right), for the LIN, PCAT, and EXT samples, with an SFH-based division into early and late galaxies. The values of $\chi^2$ are from jointly modelling $[\wproj, \xi_0, \xi_2]$. For clarity, vertical offsets are applied to data points and model curves, as indicated in the upper panels. Ratios of early-to-late galaxy measurements and best model fits are in the lower panels. The format is similar to that in  Fig.~\ref{fig:SFHwp}.}
\label{fig:SFHrsd}
\end{figure*}
%%%%%%%%%%%%%%%%%%%%%%%%%%%%%%%%%%%%%%%%%%%%%

%%%%%%%%%%%%%%%%%%%%%%%%%%%%%%%%%%%%%%%%%%%%%
\begin{figure*}
\includegraphics[width=\textwidth]{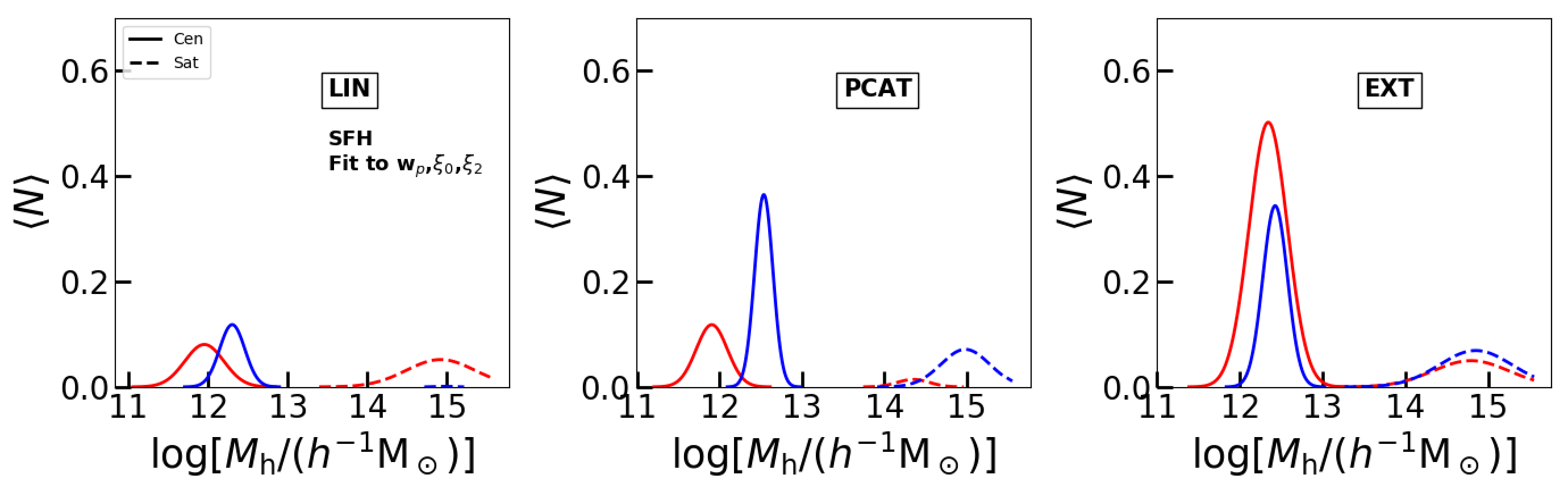}
\caption{
Mean occupation functions from models best-fitting the $\xi=[\wproj, \xi_0, \xi_2]$ measurements for the SFH-based LIN, PCAT, and EXT samples, separated into contributions from central galaxies (solid curves) and satellite galaxies (dashed curves).
}
\label{fig:SFHhod}
\end{figure*}
%%%%%%%%%%%%%%%%%%%%%%%%%%%%%%%%%%%%%%%%%%%%%

%%%%%%%%%%%%%%%%%%%%%%%%%%%%%%%%%%%%%%%%%%%%%
\begin{figure*}
\includegraphics[width=\textwidth]{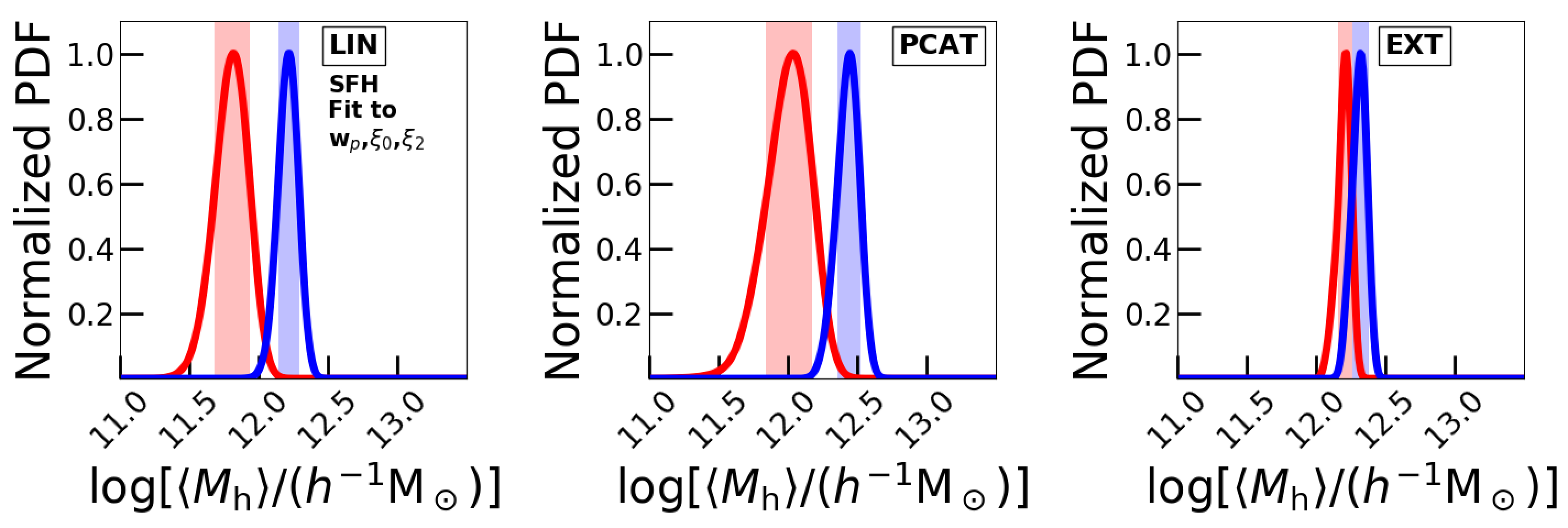}
\caption{
Same as Fig.~\ref{fig:SFHmassCONwp}, but from modelling the projected 2PCFs and redshift-space multipoles, $\xi=[\wproj, \xi_0, \xi_2]$ for the LIN, PCAT, and EXT samples (from left to right), with an SFH-based division into early and late galaxies. 
To characterise the difference between the distributions for early and late galaxies, we compute the significance values of the split between the peaks, which are 2.94$\sigma$ (LIN), 2.46$\sigma$ (PCAT), and 1.39$\sigma$ (EXT). The mean mass constraints are listed in Table~\ref{table:gals}.
}
\label{fig:SFHmassCONrsd}
\end{figure*}
%%%%%%%%%%%%%%%%%%%%%%%%%%%%%%%%%%%%%%%%%%%%%%%%

%%%%%%%%%%%%%%%%%%%%%%%%%%%%%%%%%%%%%%%%%%%
\begin{figure*}
\includegraphics[width=\textwidth]{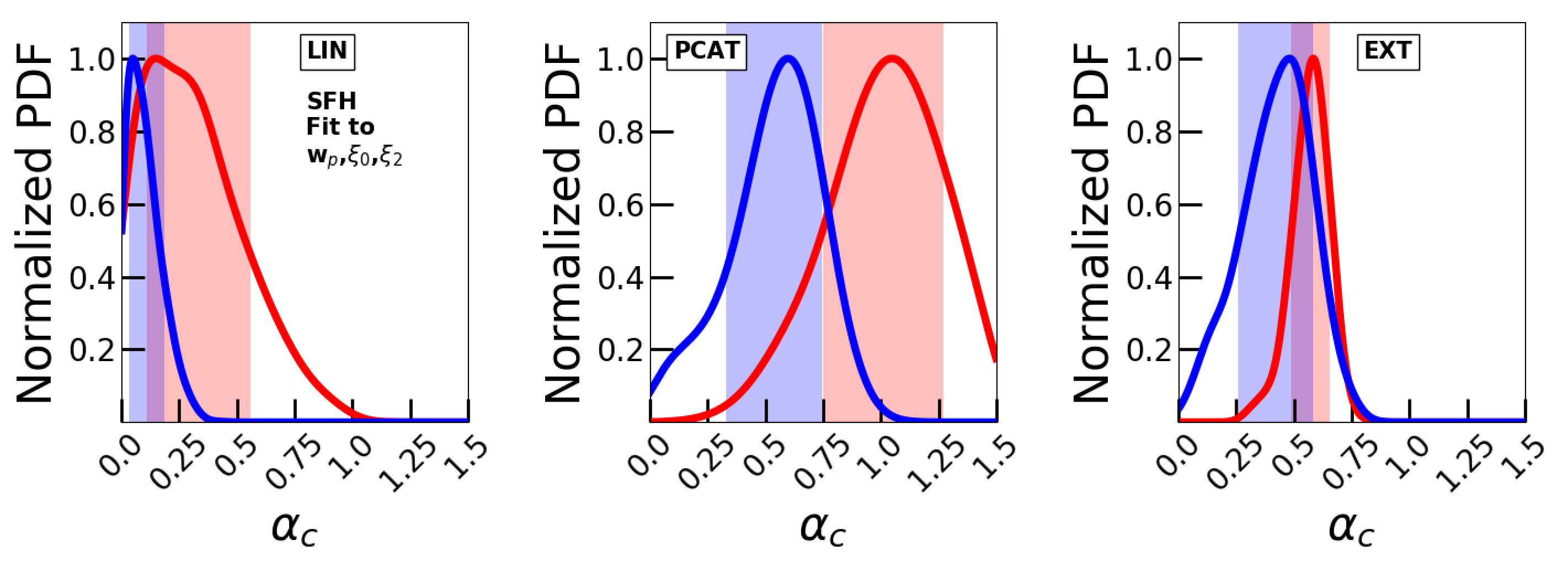}
\caption{
Constraints central galaxy velocity bias $\alphac$ for the SFH-based LIN, PCAT, and EXT samples (from left to right).
For each sample, the curve show the PDF of the constraints, with the shaded region marking the central 68.3\% distribution. 
 The constraints on $\alpha_{\rm c}$ for early and late galaxy samples are follows: 
 $0.150^{+0.408}_{-0.041}$ and $0.050^{+0.135}_{-0.015}$ (LIN); 
 $1.05^{+0.22}_{-0.30}$ and $0.60^{+0.15}_{-0.27}$ (PCAT); 
 $0.583^{+0.071}_{-0.096}$ and $0.48^{+0.101}_{-0.221}$ (EXT).
To characterise the difference between the distributions for early and late galaxies, we compute the significance values of the split between the peaks, which are 0.71$\sigma$ (LIN), 1.35$\sigma$ (PCAT), and 0.74$\sigma$ (EXT). 
}
\label{fig:SFHalphacCON}
\end{figure*}
%%%%%%%%%%%%%%%%%%%%%%%%%%%%%%%%%%%%%%%%%%%%%

%%%%%%%%%%%%%%%%%%%%%%%%%%%%%%%%%%%%%%%%%%%%%
\begin{figure*}
\includegraphics[width=\textwidth]{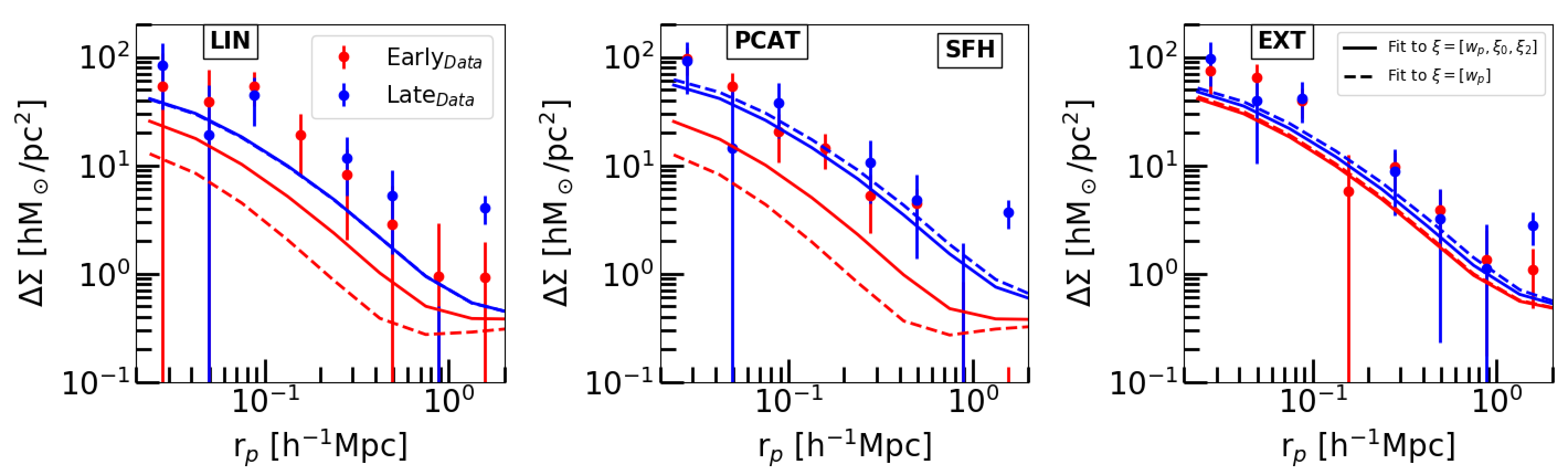}
\caption{
ESD from lensing measurements (data points) and from best-fitting model predictions (curves) for the SFH-based LIN, PCAT, and EXT samples (from left to right). Predictions from fits to $\xi=[\wproj]$ and $\xi=[\wproj,\xi_0,\xi_2]$ are shown as solid and dashed curves, respectively. See discussions in the text.}
\label{fig:SFHesd}
\end{figure*}
%%%%%%%%%%%%%%%%%%%%%%%%%%%%%%%%%%%%%%%%%%%%%%%%

%%%%%%%%%%%%%%%%%%%%%%%%%%%%%%%%%%%%%%%%%%%%%%%%%%%%%%%%%%%%%%%%%%%%%%%%%%%%%%%%%%%%%%%%%%%%%%%%%%%%%%
\section{Results}
\label{sec:Results}

In this section, we present the modelling results for the SFH-based and sSFR-based samples, respectively. For each of the LIN, PCAT, and EXT catalogue constructions (section~\ref{sec:data} and Table~\ref{table:gals}), the early and late samples are jointly modelled. To reduce any residual effect caused by fibre collision correction, we limit the modelling to scales above $2\hinvMpc$.

We first consider the case of modelling only the projected 2PCFs $\xi=[\wproj]$ (and number densities). In this case, the free parameters do not include the velocity bias parameters (with $\alphac$ and $\alphas$ fixed at 0 and 1, respectively). We focus on the constraints on the mean mass of the haloes hosting central galaxies to study the assembly bias signature.

We then model both the projected 2PCFs and redshift-space 2PCF multipoles, $\xi=[\wproj,\xi_0,\xi_2]$, together with the number densities. We focus on presenting the results on both the mean mass of central host haloes and the central velocity bias parameter $\alphac$.

We also show the consistency check with the weak lensing measurements by comparing the ESD measured from the data and predicted from the best-fitting models. 

\subsection{Results for SFH-based Samples}
\label{SFHres}

We start with the early/late samples split based on SFH. Fig.~\ref{fig:SFHwp} shows the projected 2PCF $\wproj$ measurements (points) and the results from modelling $\wproj$ only (curves for bestfit and shaded regions for the 1--$\sigma$ ranges), with the LIN, PCAT, and EXT samples (left to right panels). We use the notation that the results for the early (late) sample are in red (blue). 

In all cases, by eye, the modelling curves are able to reproduces the measurements. However, the values of $\chi^2$ appear to be large, given the degrees of freedom (dof). For each set of early/late samples, there are 14 data points (12 for $\wproj$ and 2 for $\ngg$) and the model has 12 free parameters (3 for the mean occupation function of central galaxies and 3 for that of satellites for the early sample, and the same numbers for the late sample; see equations~\ref{eq:Ncen} and \ref{eq:Nsat}). We then have dof=2. The large values of $\chi^2$ could result from the noise in the estimated covariance matrix and/or the neglect of the correlation between the measurements of early and late samples.

In each case, the amplitude of $\wproj$ of the late sample appears to be slightly higher than that of the early sample. Their ratio (bottom panels of Fig.~\ref{fig:SFHwp}) are consistent with unity, given the uncertainty. The left panels simply reproduce the L16 results that no clear evidence is found for the assembly bias with the early and late galaxies (with the measurements improved with fibre collision correction). As discussed in the sample construction, the LIN sample removed some central galaxies (mislabelled as satellites), especially for those in pairs having similar lines of sight. This results in the substantial drop of projected 2PCFs towards smaller scales (e.g. below 1$\hinvMpc$). The PCAT and EXT samples have recovered those mislabelled central galaxies, and the corresponding projected 2PCFs do not show the drop towards small scales. With a fraction of central-central pairs along similar lines of sight missing, the redshift-space clustering for LIN samples would become hard to interpret and model.  Hereafter, we show the modelling results with LIN samples for complete purposes and will not attempt to draw conclusions from them. We will focus on presenting the results from the PCAT and EXT samples.

The PCAT and EXT samples specifically include satellites by construction, and the modelling results show a low satellite fraction. To see whether the small difference in the $\wproj$ amplitude signals assembly bias effect, we turn to the constraints on the mean mass of haloes hosting central galaxies.

The probability distribution functions (PDF) of the mean halo mass for central galaxies are shown in Fig.~\ref{fig:SFHmassCONwp}. For the LIN samples, the 1$\sigma$ ranges of the mean host halo masses for central early/late galaxies overlap, and the values are consistent with those inferred from lensing measurements in L16 ($\log\langle \Mh/(\hinvMsun) \rangle = 11.98^{+0.14}_{-0.10}$ and $11.92^{+0.10}_{-0.10}$, respectively). For the PCAT samples, the difference between the mean halo masses of central early/late galaxies becomes larger, although still being consistent. For the EXT samples, the mean halo mass PDFs overlap and the constraints become tighter, resulting from the larger samples.

Similar to the projected 2PCFs, for redshift-space clustering, the measurements of the 2PCF monopoles (middle panel of Fig.~\ref{fig:SFHrsd}) for early galaxies appear to be lower than those for late galaxies, while being consistent within the uncertainties. The quadrupoles of the two samples are similar (right panel of Fig.~\ref{fig:SFHrsd}). Based on the values of $\chi^2$ (in the left panels), in each case (LIN, PCAT, or PCAT), the model provides a good fit to [$w_{\rm p}$, $\xi_0$, $\xi_2$] measurements.

Fig.~\ref{fig:SFHhod} shows the mean occupation functions from the best-fitting models, separated into contributions from central galaxies (solid curves) and satellite galaxies (dashed curves). Satellite galaxies appear to to occupy haloes of mass $\gtrsim 10^{14}\hinvMsun$. In the sample construction, these satellites are misidentified as central galaxies with similar stellar masses, while the true central galaxies reside in haloes of mass $\sim10^{12} \hinvMsun$. As the halo mass function drops steeply towards the high mass end, the fraction of galaxies being satellites is low (as shown in the middle panels of Fig.~\ref{fig:SFHrsd}).

As shown in Fig.~\ref{fig:SFHmassCONrsd}, the trend in the constraints on the mean halo mass of central galaxies is similar to that with the projected 2PCFs, with halo mass for late galaxies being generally more massive. The addition of redshift-space 2PCF multipoles have produced tighter constraints, and the split between mass scales of early and late central galaxies becomes more significant (with $\sim 2.5\sigma$ and $\sim 1.8\sigma$ for the PCAT and EXT samples).

In Fig.~\ref{fig:SFHalphacCON} the constraints on the central-galaxy velocity bias parameter ($\alphac$) are compared. As mentioned before, because the LIN samples miss a fraction of central-central galaxy pairs along similar lines of sight, we do not consider redshift-space constraints from the LIN samples to be informative and simply show them for consistency in the presentation of the results. The $\alphac$ constraints for early and late central galaxies are consistent for both the PCAT and EXT samples, with those for PCAT sample having larger uncertainties. While there is a trend that early central galaxies have higher velocity bias (with lower mean halo mass), the large uncertainties prevent a definitive conclusion. 

The apparent high values of central velocity bias in Fig.~\ref{fig:SFHalphacCON} appear surprising. For the PCAT and EXT samples, the PDFs of $\alphac$ peak around 0.5 or above. This is different from the central velocity bias constraints with luminosity-threshold samples \cite[][]{Guo15c}, which is usually lower than $\alphac=0.5$ with median values of 0.2--0.3. In the galaxy formation model investigated by \citet{Ye17}, high $\alphac$ is possible but only for central galaxies with low stellar mass to halo mass ratio. However, if we account for the full distribution of $\alphac$, $\alphac=0.2$--0.3 falls into the 3$\sigma$ range of the constraints. One possible cause of the tendency of high $\alphac$ is that some satellite galaxies (with higher velocity dispersion) are misidentified as central galaxies. However, we do include satellites in our model, which is supposed to capture the satellite contribution. In the parameterisation, the mean occupation function of satellites assumes a log-normal form (equation~\ref{eq:Nsat}). We perform tests by changing it to be a power-law form, and the resultant $\alphac$ constraints remain similar. The other possibility is that the tendency of high $\alphac$ may be some manifestation of the assembly bias effect (see discussion in section~\ref{sec:discussion}). This possibility needs to be addressed with more sophisticated models. But given the uncertainties seen in the constraints here, the difference in the assembly bias effects of the early and late central galaxies are unlikely to be established with high statistical significance.

In Fig.~\ref{fig:SFHesd}, we show the ESD measurements from gravitation lensing on scales below 2$\hinvMpc$. In each case, at $\rp\sim 1.5\hinvMpc$, the ESD around late galaxies appears to be higher than that around early galaxies. It is not clear whether this is a statistical fluctuation or not. Except for this point, the ESD measurements around early and late galaxies agree with each other, which indicates similar mean host halo mass for populations dominated by central galaxies. The curves show the predictions from mocks generated by populating haloes in the {\sc MultiDark MDPL2} $N$-body simulation from the HOD that best-fits $\wproj$ (dashed) and [$\wproj$, $\xi_0$, $\xi_2$] (solid).

The model from fitting $\wproj$ appears to underpredict the ESD for early galaxies for the LIN and PCAT samples. If the uncertainties in the halo mass constraints  (Fig.~\ref{fig:SFHmassCONwp}) are taken into account, the discrepancy can be reduced. In other words, if the ESD were also used to constrain the model, the model would be likely to shift to better agree with the ESD measurements.

The underprediction also shows up when fitting [$\wproj$, $\xi_0$, $\xi_2$], but to a less extent. As mentioned above, the modelling results of the LIN sample are hard to interpret and we focus on the PCAT sample.
For the PCAT case, the model reproduces the ESD for late galaxies, but underpredicts that for early galaxies. The underprediction is in line with the lower mean halo mass of early galaxies, as shown in the middle panel of Fig.~\ref{fig:SFHmassCONrsd}. Again the discrepancy can be reduced when accounting for the uncertainties in halo mass constraints. On the other hand, the results of low halo mass, high $\alphac$, and low ESD for the PCAT early galaxies, if taken together, may hint assembly bias effect -- for example, these early central galaxies may reside in haloes of higher mass but late-formed (thus lower clustering amplitude); therefore, the model (without assembly bias) puts them in lower mass haloes (thus lower ESD); the resultant lower velocity dispersion inside haloes requires a higher $\alphac$ to fit the redshift-space clustering. Further work is needed to explore this possibility.

For the EXT case, the model ESD curves agree with the ESD measurements for both late and early galaxies. The predicted ESD for early galaxies is slightly lower. The overall trend in the results of mean halo mass, $\alphac$, and ESD for early galaxies is similar to the PCAT case but much weaker. 

As a whole, with the SFH-based early and late samples with different constructions (PCAT and EXT), we find that the HOD modelling results support that the central galaxies reside in haloes of similar mass. The differences seen in mean halo mass and central velocity bias parameter may be attributed to assembly bias effect, but the differences in the constraints are not large enough to make definitive conclusions. With a model incorporating assembly bias descriptions, we would be able to more quantitatively assess the assembly bias effect by jointly fitting the projected and redshift-space 2PCFs and the ESD.

%%%%%%%%%%%%%%%%%%%%%%%%%%%%%%%%%%%%%%%%%%%%%
\begin{figure*}
\includegraphics[width=\textwidth]{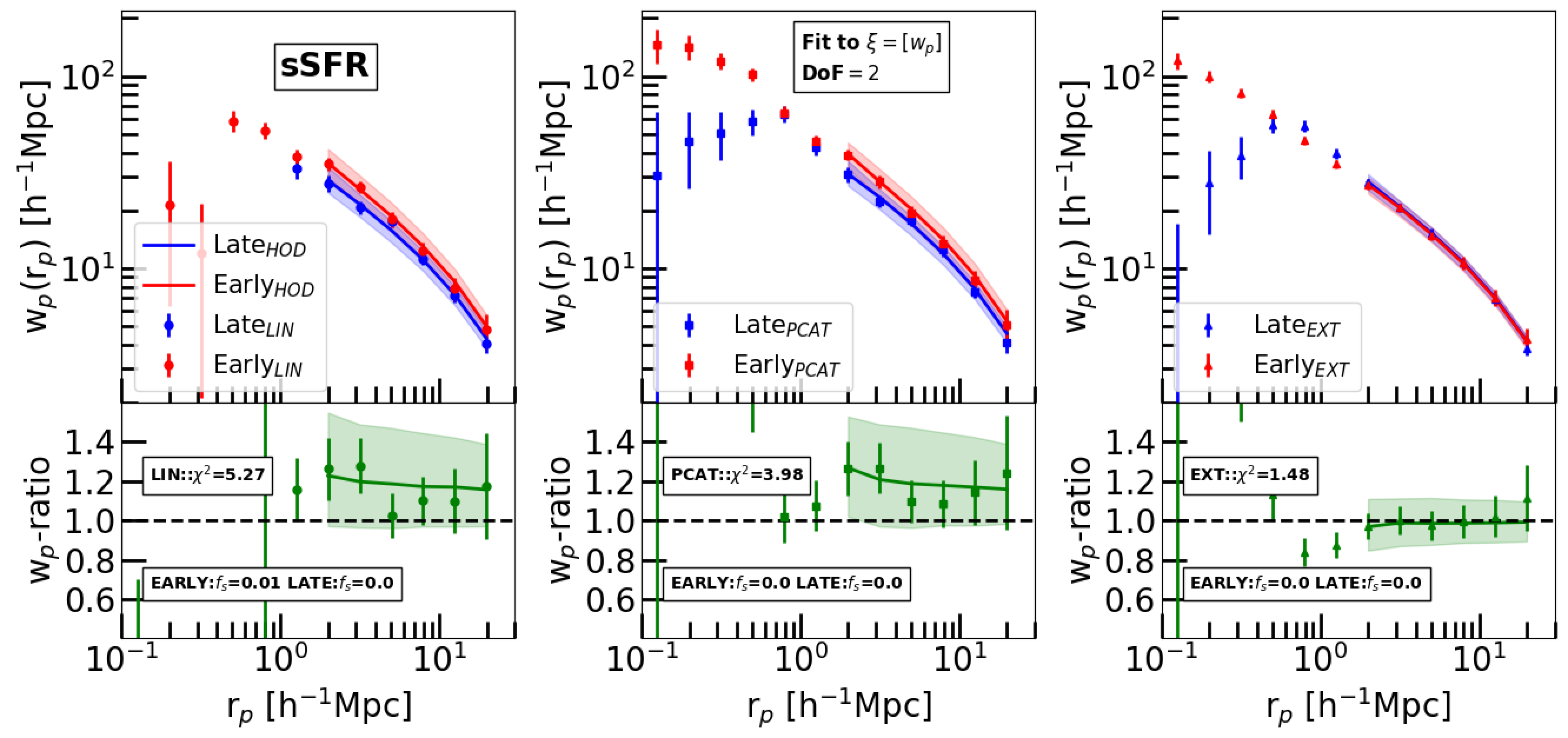}
\caption{
Same as Fig.~\ref{fig:SFHwp}, but with an sSFR-based division into early and late galaxies.
}
\label{fig:sSFRwp}
\end{figure*}
%%%%%%%%%%%%%%%%%%%%%%%%%%%%%%%%%%%%%%%%%%%%%%%%

%%%%%%%%%%%%%%%%%%%%%%%%%%%%%%%%%%%%%%%%%%%%%
\begin{figure*}
\includegraphics[width=\textwidth]{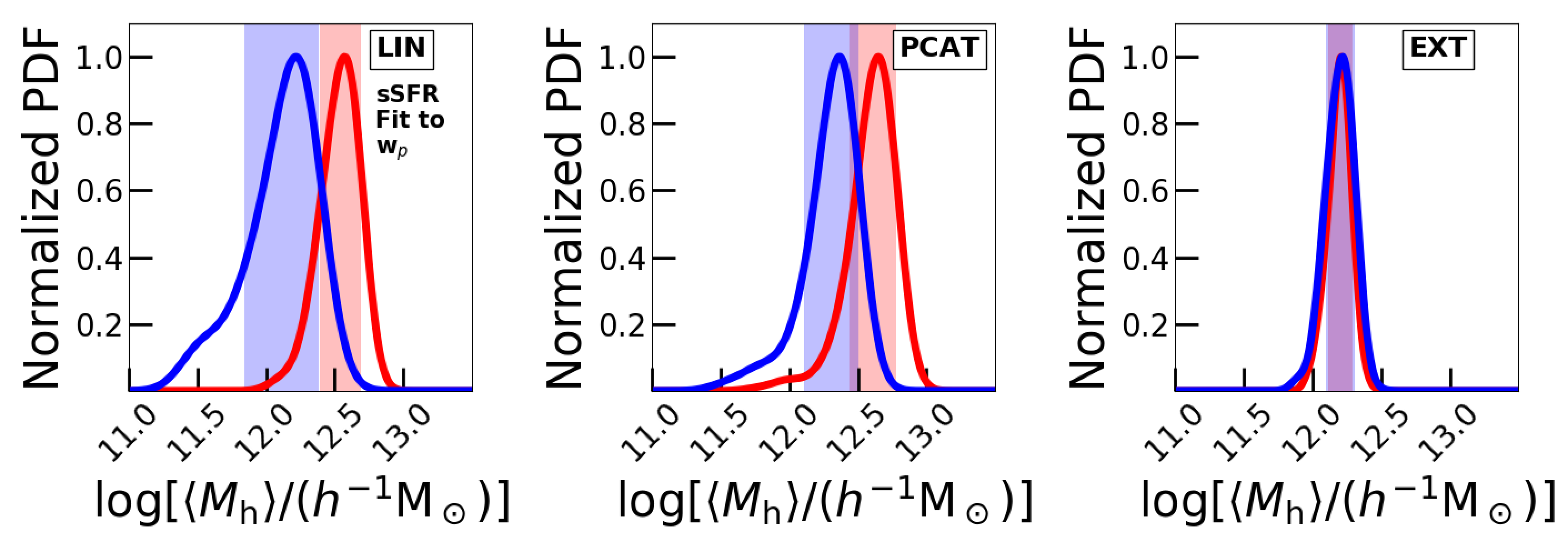}
\caption{
Same as Fig.~\ref{fig:SFHmassCONwp}, but with an sSFR-based division into early and late galaxies. 
To characterise the difference between the distributions for early and late galaxies, we compute the significance values of the split between the peaks, which are 1.45$\sigma$ (LIN), 1.15$\sigma$ (PCAT), and 0.07$\sigma$ (EXT). The mean mass constraints are listed in Table~\ref{table:gals}.
}
\label{fig:sSFRmassCONwp}
\end{figure*}
%%%%%%%%%%%%%%%%%%%%%%%%%%%%%%%%%%%%%%%%%%%%%%%%

\subsection{Results for sSFR-based Samples}
\label{sSFRres}

%%%%%%%%%%%%%%%%%%%%%%%%%%%%%%%%%%%%%%%%%%%%%
\begin{figure*}
\includegraphics[width=\textwidth]{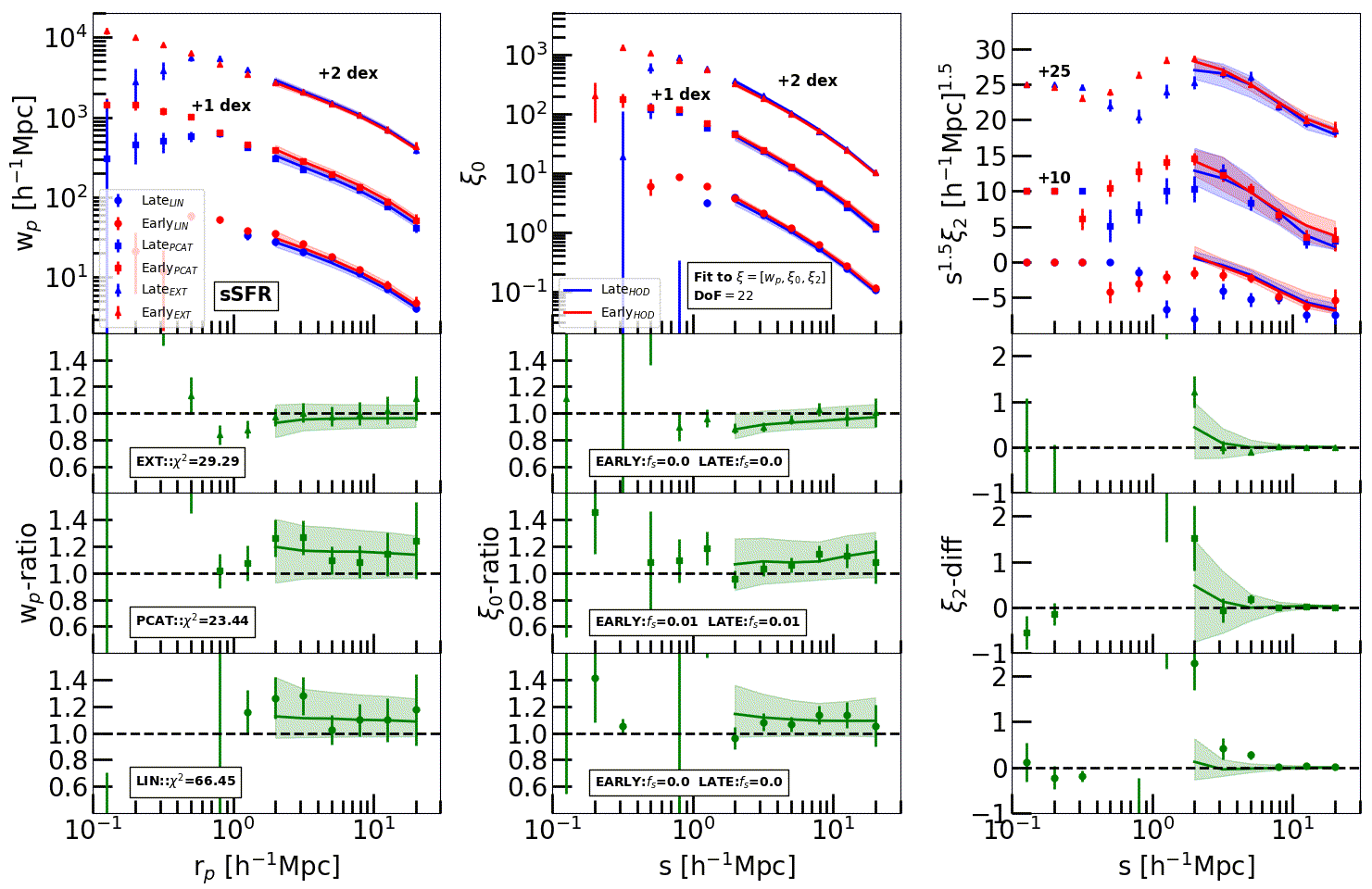}
\caption{
Same as Fig.~\ref{fig:SFHrsd}, but with an sSFR-based division into early and late galaxies. 
}
\label{fig:sSFRrsd}
\end{figure*}
%%%%%%%%%%%%%%%%%%%%%%%%%%%%%%%%%%%%%%%%%%%%%%%%

%%%%%%%%%%%%%%%%%%%%%%%%%%%%%%%%%%%%%%%%%%%%%
\begin{figure*}
\includegraphics[width=\textwidth]{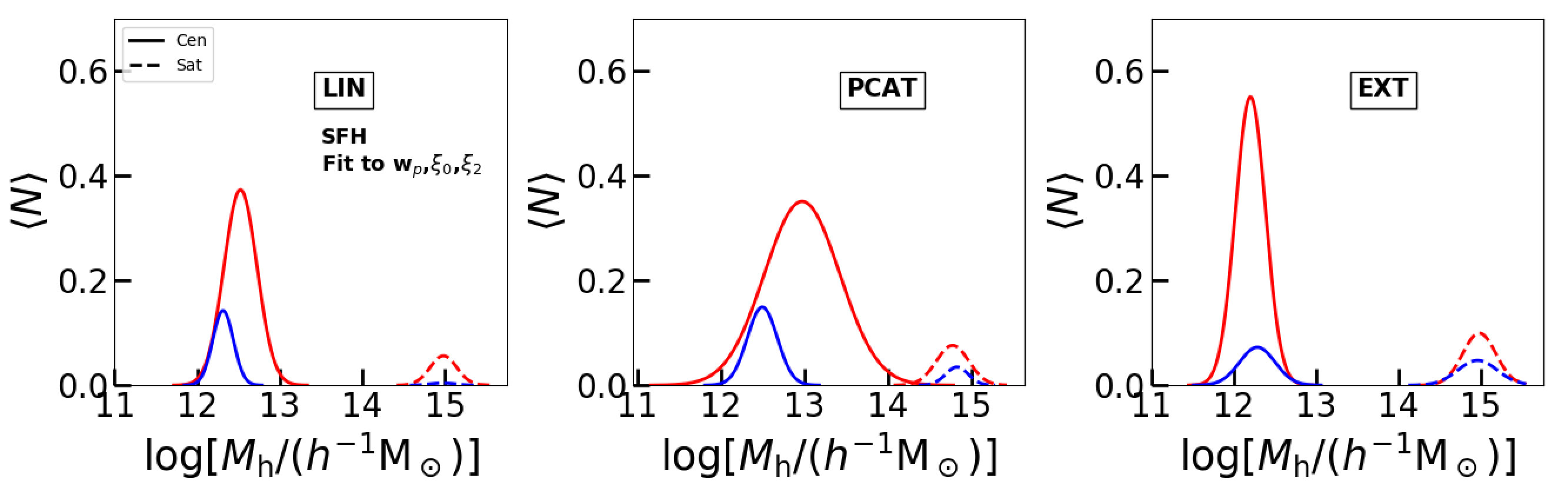}
\caption{Same as Fig.~\ref{fig:SFHhod}, but for the sSFR-based samples.
}
\label{fig:sSFRhod}
\end{figure*}
%%%%%%%%%%%%%%%%%%%%%%%%%%%%%%%%%%%%%%%%%%%%%

%%%%%%%%%%%%%%%%%%%%%%%%%%%%%%%%%%%%%%%%%%%%%
\begin{figure*}
\includegraphics[width=\textwidth]{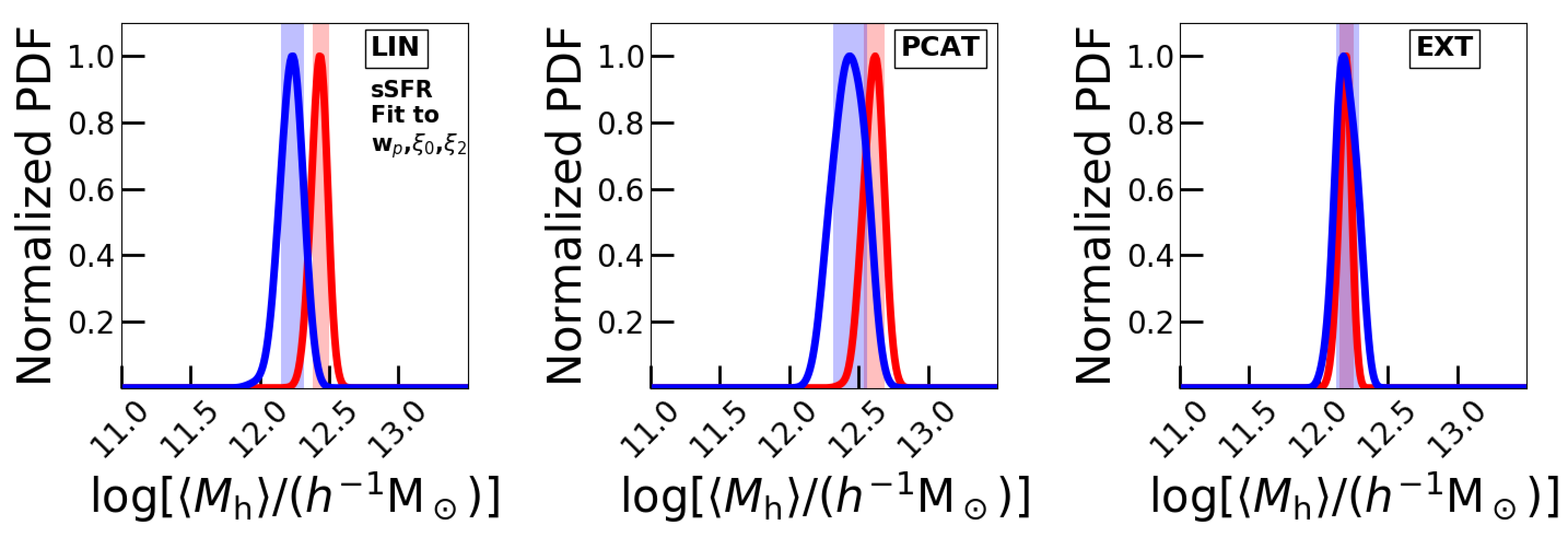}
\caption{
Same as Fig.~\ref{fig:SFHmassCONrsd}, but with an sSFR-based division into early and late galaxies. To characterise the difference between the distributions for early and late galaxies, we compute the significance values of the split between the peaks, which are 2.01$\sigma$ (LIN), 1.25$\sigma$ (PCAT), and 0.17$\sigma$ (EXT). The mean mass constraints are listed in Table~\ref{table:gals}.
}
\label{fig:sSFRmassCONrsd}
\end{figure*}
%%%%%%%%%%%%%%%%%%%%%%%%%%%%%%%%%%%%%%%%%%%%%%%%

%%%%%%%%%%%%%%%%%%%%%%%%%%%%%%%%%%%%%%%%%%%%%
\begin{figure*}
\includegraphics[width=\textwidth]{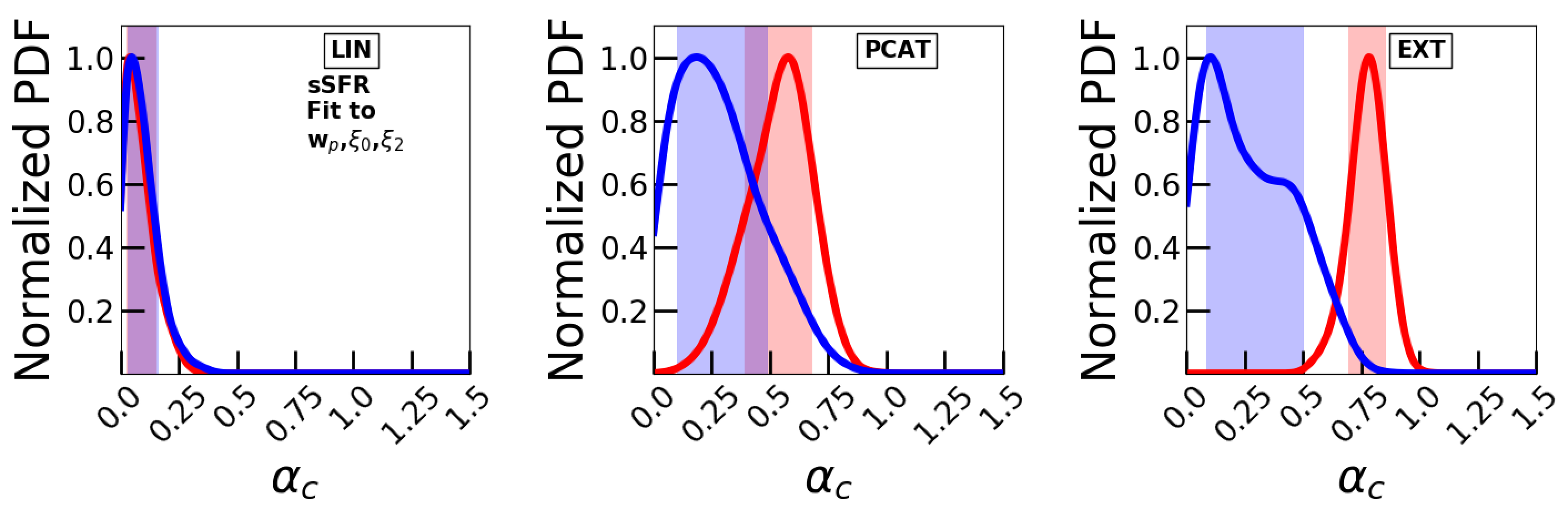}
\caption{
Same as Fig.~\ref{fig:SFHalphacCON}, but with an sSFR-based division into early and late galaxies. The constraints on $\alpha_{\rm c}$ for early and late galaxy samples are follows: $0.040_{-0.012}^{+0.115}$ and $0.048_{-0.016}^{+0.017}$ (LIN); $0.58_{-0.19}^{+0.10}$ and $0.185^{+0.306}_{-0.082}$ (PCAT); $0.784^{+0.076}_{-0.093}$ and $0.103^{+0.401}_{-0.017}$ (EXT).
To characterise the difference between the distributions for early and late galaxies, we compute the significance values of the split between the peaks, which are 0.07$\sigma$ (LIN), 1.09$\sigma$ (PCAT), and 1.66$\sigma$ (EXT). 
}
\label{fig:sSFRalphacCON}
\end{figure*}
%%%%%%%%%%%%%%%%%%%%%%%%%%%%%%%%%%%%%%%%%%%%%%%%

%%%%%%%%%%%%%%%%%%%%%%%%%%%%%%%%%%%%%%%%%%%%%
\begin{figure*}
\includegraphics[width=\textwidth]{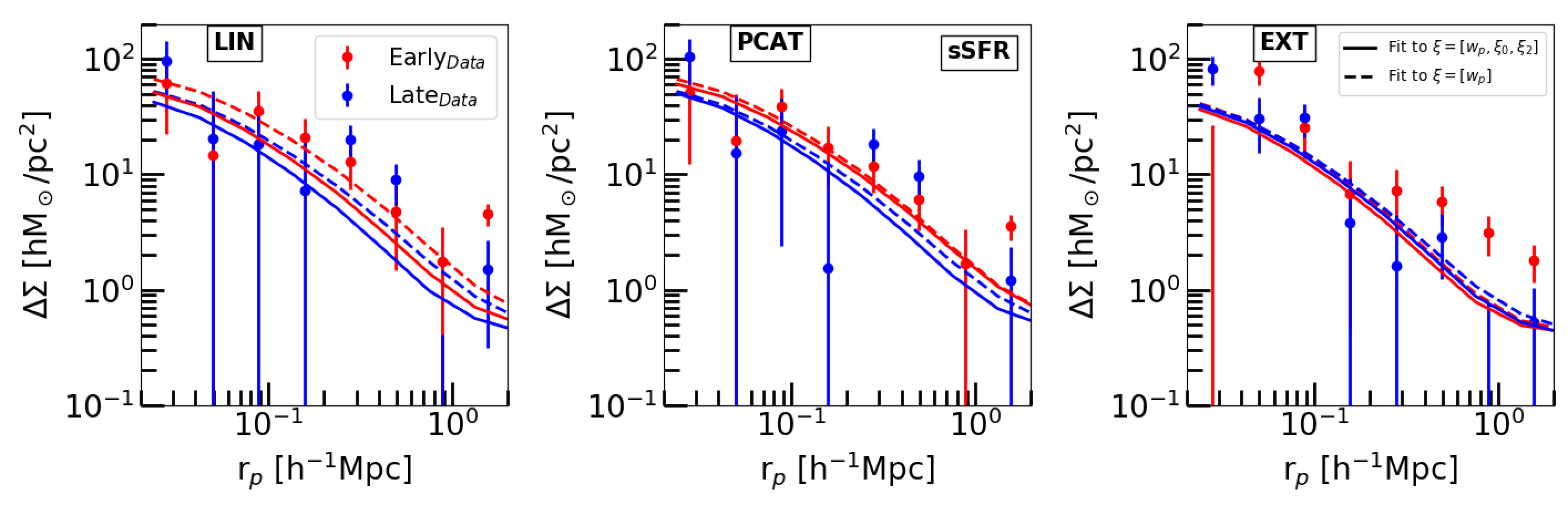}
\caption{
Same as Fig.~\ref{fig:SFHesd}, but with an sSFR-based division into early and late galaxies. 
}
\label{fig:sSFResd}
\end{figure*}
%%%%%%%%%%%%%%%%%%%%%%%%%%%%%%%%%%%%%%%%%%%%%

The $\wproj$ measurements and the corresponding model fits for early/late samples constructed according to sSFR are shown in Fig.~\ref{fig:sSFRwp}. Opposite to the SFH-based samples, the projected 2PCFs for early galaxies in the LIN and PCAT cases have higher clustering amplitudes than late galaxies, while the amplitudes are similar in the EXT case (on scales above $2\hinvMpc$). However, the amplitude ratio in each case is consistent with unity (particularly so in the EXT case), given the uncertainties. The HOD model provides good fits to the measurements in all the cases.

Compared to the SFH-based samples, the mean halo mass for central galaxies is better constrained (Fig.~\ref{fig:sSFRmassCONwp}). While the halo mass scale for early central galaxies appears to be slightly higher than that for late central galaxies, the central $1\sigma$ constraints for the mass scales of early and late central galaxies overlap with each other. In particular, the mass scale constraints for early/late galaxies in the EXT case fall on top of each other, a manifestation of the similar amplitudes in $\wproj$ (Fig.~\ref{fig:sSFRwp}). In the LIN case, the mass scales are consistent with those from lensing measurements in L16 ($\log\langle \Mh/(\hinvMsun) \rangle=12.14\pm 0.07$ and $12.10\pm 0.08$ for early and late central galaxies).

Fig.~\ref{fig:sSFRrsd} shows the measurements of both the projected and redshift-space 2PCFs, $\xi$=[$\wproj$, $\xi_0$, $\xi_2$], and the fitting results. For the same reason as with the SFH-based samples, we focus on the PCAT and EXT cases. We note that for both the PCAT and EXT late galaxy samples, the quadrupoles below $2\hinvMpc$ become negative, opposite to the expected Finger-of-God effect. We perform tests and discuss the possible origin in section~\ref{sec:discussion}. On scales above $2\hinvMpc$, the amplitudes of the projected and redshift-space 2PCFs for early and late galaxies agree with each other. The HOD model results in reasonable values of $\chi^2$ when fitting jointly to $\xi$=[$\wproj$, $\xi_0$, $\xi_2$]. The mean occupation functions from the best-fitting models are shown in Fig.~\ref{fig:sSFRhod}, with trend similar to that in Fig.~\ref{fig:SFHhod}.

The halo mass constraints (Fig.~\ref{fig:sSFRmassCONrsd}) from jointly modelling $\xi$=[$\wproj$, $\xi_0$, $\xi_2$] are similar to those from modelling $\wproj$ only (Fig.~\ref{fig:sSFRmassCONwp}), but becoming much tighter with the benefit of the redshift-space clustering. 

For the constraints on the central velocity bias (Fig.~\ref{fig:sSFRalphacCON}), similar to the trend seen in the SFH-based samples, both the PCAT and EXT cases have a higher $\alphac$ for early galaxies. The 1$\sigma$ range of $\alphac$ for the two cases are (0.39, 0.69) and (0.70, 0.86), respectively. We note that the velocity bias for the EXT early galaxy case is substantially higher than that for luminosity-threshold samples \citep[][]{Guo15c}.

For the sSFR-based samples, the ESD measurements and the predictions from the best-fitting HOD models are shown in  Fig.~\ref{fig:sSFResd}. The broad agreement between the measurements and predictions lends further support that the samples constructed have similar host halo masses for central galaxies. For the EXT case, the measured ESD around early galaxies appears to be higher than the predictions. In combination with the high $\alphac$ for the EXT early galaxies, it may point toward a possibility that a fraction of satellites in high mass haloes are misidentified as central galaxies in haloes of $\sim 10^{12}\hinvMsun$. However, our model includes contributions from such satellites, and like the SFH-based case our tests with an alternative parameterisation for satellite occupation function does not change the results. Assembly bias remains as a possible explanation, which will be discussed in section~\ref{sec:discussion}.

As a whole, with the sSFR-based samples, we obtain results similar to those with the SFH-based samples. Although there are overall agreement between the modelling results of early and late central galaxies, early central galaxies tend to have higher values of velocity bias.

\section{Discussion}
\label{sec:discussion}

The main model constraints we are interested in this work are the mass scales of central galaxy host haloes and the velocity bias of central galaxies. Unlike the samples used in \citet{Lin16}, called LIN samples in this paper, we do not perform the satellite decontamination algorithm in defining the extended PCAT and EXT samples in order to study the redshift-space clustering. While we have parameters for satellite components in our HOD model, for the purpose of studying central galaxies, they are treated as nuisance parameters to absorb the effect of satellites. We focus our discussion on the results with the PCAT and EXT samples.

For the PCAT samples, the mass scales of central hosting haloes of early and late galaxies show a split. While in the SFH-based samples early galaxies tend to reside in haloes of lower mass than late galaxies, the trend is opposite in the sSFR-based samples. However, the split in either SFH-based or sSFR-based samples is not statistically significant from the marginalised distributions of the mass scale. Compared with the constraints from $\wproj$ measurements only, including redshift-space multipoles reduces the uncertainty in the mass scale by a factor of about two, and the split remains insignificant. 

When constructing the EXT samples, we not only exclude the satellite decontamination procedure but also adjust the range of galaxy properties to substantially extend the samples. As a consequence, the EXT samples lead to tight constraints on the mass scale of central hosting haloes, and adding the redshift-space clustering further tightens the constraints. In our full analyses with projected and redshift-space clustering, the uncertainties in the mean mass scale are about 0.05 dex for either the SFH-based or the sSFR-based samples. Even with such tight constraints, no significant difference between the mass scales of early and late central galaxies are found.

As a consistency check, we compare the HOD modelling results to the lensing measurements. In general, the predicted ESD values measured from mock galaxy catalogues based on the best-fitting HOD show a broad agreement with those measured from the data. The main exception is with the SFH-based early galaxy sample. The predicted ESD curves from the $\wproj$-only fit are about 1 dex lower than the measurements. This is in line with the difference in the halo mass constraints between early and late galaxies. Accounting for the large uncertainty ($\sim$0.33 dex) in the mass scale constraints reduces the discrepancy. With the inclusion of redshift-space clustering, the predicted ESD curve for early galaxies is shifted upward by $\sim$0.5 dex. It is likely that the apparent mismatch between the predicted and measured ESD for the SFH-based early galaxy sample reflects the loose constraint on the halo mass scale and that the current best-fitting model for the SFH-based early galaxies is somewhat biased. Including ESD in the data vector in future investigations will tighten model constraints and assist in selecting the appropriate models.

The most important results of this work are the constraints on the central galaxy velocity bias ($\alphac$), shown in Fig.~\ref{fig:SFHalphacCON} and Fig.~\ref{fig:sSFRalphacCON} for the SFH-based and sSFR-based samples, respectively. Again our focus here are on the PCAT and EXT sample constructions (middle and right panels). There appears to be a split between the values of central velocity bias of early and late samples, in the sense of higher velocity bias for early galaxies. For the SFH-based samples, one may associate this with the apparent split in halo mass scale (fig.~\ref{fig:SFHmassCONrsd}) and argue for a halo mass dependent velocity bias. However, the trend of higher $\alphac$ in lower mass haloes appears to be opposite to that inferred from luminosity-threshold samples \citep[][]{Guo15c}. For the sSFR-based samples, while there is a split in $\alphac$, there is no clear split in the halo mass scale. Being unlikely attributed to the halo-mass dependence, the split in $\alphac$ of early and late galaxies  may signal a galaxy assembly bias effect. In the model, the central velocity bias characterises the mutual motion (relaxation) between central galaxies and their host haloes. If early galaxies reside in early-formed haloes \citep[][]{Xu20}, the split would indicate that galaxies in early-formed haloes are less relaxed, contrary to the naive expectation. As investigated in \citet{Ye17} with the Illustris simulation, the magnitude of velocity bias is related to the history of major and minor mergers, and the above split would provide information on the merging history of haloes and galaxies. However, the constraints on central velocity bias are loose and the split is not statistically significant, with the biggest split being about 1.7$\sigma$ in the sSFR-based EXT samples. Taking the face values, we find no evidence of significant difference in the central velocity bias of early and late galaxies. Then we would reach the conclusion that no assembly bias effect is detected in the motion of central galaxies relative to their dark matter haloes.

However, the surprise lies in the value of central velocity bias. If we use the central velocity bias constraints from luminosity-threshold samples \citep[][]{Guo15c} as a comparison point, the constraints on the central velocity bias for late galaxies in this work are in the ballpark of those for luminosity-threshold samples. However, those for early galaxies are much higher, roughly lying in the range of $\alphac\sim$ 0.5--1.0. That is, in the halo rest frame early central galaxies would on average move at a velocity on the order of 35--70\% that of the halo virial velocity. The stellar mass to halo mass ratio of these central galaxies is around 1 per cent (section~\ref{sec:data}), and in the theoretical investigation of \citet{Ye17}, the expected velocity bias is about 0.32 (see their fig.2). Could this high velocity bias result from mistaking satellites as central galaxies? First, we account for satellite galaxies in the model, and we find similar central velocity bias constraints with different parameterisations of the mean satellite occupation function. Second, given the selection criteria, galaxies in the samples are dominated by central galaxies in haloes around $10^{12}\hinvMsun$ and their stellar mass is a few times $10^{10}h^{-2}\Msun$ (section~\ref{sec:data}). It would be rare for such haloes to host such massive satellites. Such satellites can reside in more massive halos, but given the low satellite fraction they would not play a role in significantly interfering with the central velocity bias constraints. Therefore, the apparent high velocity bias inferred for early central galaxies is unlikely caused by mistaking satellites as central galaxies. 

The high value of velocity bias for early central galaxies may be a signal of assembly bias effect, but here the inferred velocity bias does not necessarily reflect the mutual motion between central galaxies and host haloes. With early and late central galaxies in the PCAT or EXT samples residing in haloes of similar mass, we make the assumption that early central galaxies on average are found in early formed haloes \citep[e.g.][]{HW13,Xu20}. Earlier formed haloes have a higher pairwise velocity dispersion -- in haloes of $\sim 10^{12}\Msun$, haloes in the quartile of earliest formation can have a pairwise velocity dispersion 1.5 $\times$ higher than the average at this mass scale at a halo pair separation of $\sim 1\hinvMpc$ (see middle panels in fig.11 of \citealt{Xu18}). In modelling the redshift-space clustering using the HOD model that neglects assembly bias effect, such a high halo pairwise velocity dispersion would be captured. But it would be interpreted as pairwise velocity between central galaxies, causing the inferred central galaxy velocity bias to be high. While the above scenario may lead to difference in the clustering amplitude, the difference depends on how strong the correlation between galaxy properties and halo assembly history and it may be buried in the measurement uncertainties if the correlation is not strong.

So the assembly bias effect in halo pairwise velocity could offer an explanation on the velocity bias constraints. In haloes of fixed mass, the spatial clustering and the pairwise velocity are correlated (see \citealt{Xu18}; e.g. their fig.12). If the results in early galaxy velocity bias suggest assembly bias effect in halo pairwise velocity, other constraints from the HOD modelling would be re-interpreted. For example, the inferred mass scales for central galaxies would be modulated by the assembly bias effect and the constraints would not reflect the true mass scale. 

There may be other signs of assembly bias in the clustering measurements. As mentioned in section~\ref{sSFRres}, the quadrupole of redshift-space 2PCFs of sSFR-based late galaxies (in both the PCAT and EXT samples) appears to be negative on scales below $2\hinvMpc$ (Fig.~\ref{fig:sSFRmassCONrsd}), opposite to the expected Finger-of-God effect. We find that this feature is not caused by fibre-collision correction. To further test the fibre collision, we divide the samples into to redshift bins and perform clustering measurements. As fibre collision has a fixed angular scale, the fibre-collision effect would show up at different comoving scales for the two redshift bins. However, we obtain consistent quadrupole measurements. The negative quadrupole could result from halo exclusion effect for samples dominated by central galaxies, but we do not see such a feature in the early samples. To see whether a simple prescription of assembly bias effect can lead to such a feature, we measure the quadrupole for the blue central galaxies in the mock galaxy catalogue constructed by \citet{HW13}, where galaxy colour is tied to halo formation time. No negative quadrupole is found on scales below $2\hinvMpc$. It remains possible that this feature is related to assembly bias, but the link between galaxy properties and halo assembly variable is not as simple as in the above model.

To summarise, based on the results of analysing the redshift-space clustering of the PCAT and EXT samples of early and late galaxies (divided either by SFH or sSFR), we suspect that there are hints of galaxy assembly bias effect. Clearly a more sophisticated model incorporating assembly bias with an effective description of galaxy properties and halo assembly history is needed to further investigate the possibility and put concrete constraints on the assembly bias effect. Given the level of uncertainties seen in our constraints (especially on $\alphac$), the assembly bias may remain subtle to be detected. Therefore, a variety of observables may be needed to reveal the assembly bias signal, including the projected and redshift-space 2PCFs, the lensing measurements, and the count-in-cell statistics \citep[e.g.][]{Wang19}.

\section{Summary and Conclusion}
\label{sec:conc}

In this work, we extend the investigations of galaxy assembly bias effect in \citet{Lin16} from real space (with projected 2PCFs) to redshift (velocity) space (with redshift-space 2PCF multipoles). To achieve such a goal, we make an extension to the galaxy sample construction and divide the sample into early- and late-forming galaxies (named as early and late galaxies in this work) according to either their SFH or their sSFR. With the projected and redshift-space clustering measurements, we carry out HOD modelling to constrain the galaxy-halo relation to see whether there is any sign of assembly bias effect. Our particular interest lies in the velocity bias parameters of central galaxies, which is supposed to characterise the mutual relaxation between central galaxies and their host haloes. Differences in the velocity bias of early and late central galaxies may signal assembly bias effect.

In \citet{Lin16}, the early and late galaxies (called LIN samples here) are constructed to consist of central galaxies and weak lensing measurements show that they have consistent mean host halo mass. In this paper, based on the HOD modelling of the $\wproj$ measurements, we find that both the SFH-based and sSFR-based LIN samples have low satellite fraction (at per cent level). Within the uncertainties, the mean masses of central hosting haloes for early and late galaxies constrained in the HOD model broadly agree with each other and are consistent with the lensing measurements in \citet[][]{Lin16}. Therefore, our results based on $\wproj$ lend support to the findings in \citet[][]{Lin16}. 

The sharp drop in the $\wproj$ measurements toward the smallest scales in the LIN samples indicates that a fraction of central-central galaxy pairs along similar lines of sight ($\rp\lesssim$ 0.5--0.8$\hinvMpc$) are removed with the satellite decontamination algorithm when constructing the LIN samples. While this has little effect on analysing $\wproj$ measurements in \citet{Lin16} and in this paper, it would affect our analyses of the redshift-space clustering. Therefore we focus our study on the redshift-space clustering with the extended PCAT and EXT samples, constructed without the satellite decontamination procedure. The PCAT samples are simply the LIN samples before performing the satellite decontamination. The EXT samples extend the range of allowed stellar mass of galaxies, leading to a factor of 1.94 and 1.43 times more galaxies in the catalogue. In addition, the sSFR-based EXT samples shift the early/late galaxy dividing cut from $\log({\rm sSFR}/{\rm yr}^{-1})\sim -11.8$ to $\log({\rm sSFR}/{\rm yr}^{-1})\sim -11.0$ to better separate the quenched and active star-forming galaxies. 

Through HOD modelling of the projected 2PCFs and redshift-space 2PCF multipoles measured with the PCAT and EXT early/late galaxies, we find that the halo mass scales of central early and late galaxies, within the uncertainties, agree with each other and with those from lensing measurements inferred in \citet{Lin16}. If there is any split in the mass scale, it is only seen in the PCAT samples (middle panels of Fig.~\ref{fig:SFHmassCONrsd} and Fig.~\ref{fig:sSFRmassCONrsd}). The trend of the split depends on the construction of the early and late galaxy samples -- for SFH-based PCAT samples, early galaxies tend to reside in lower mass haloes, with the significance of the split being $\sim 2.5\sigma$; for sSFR-based PCAT samples, early galaxies tend to be in higher mass haloes, with the significance of the split being $\sim 1.2\sigma$. For each galaxy sample, we also perform lensing measurements for consistency check, and the ESD measurements largely agree with the predictions from the best-fitting HOD model.

The central galaxy velocity bias constraints from modelling both the projected and the redshift-space clustering measurements appear to be puzzling. In either the PCAT or the EXT samples, early central galaxies tend to have a higher velocity bias $\alphac$ than late galaxies. A naive explanation would be that in haloes of similar mass, the mutual relaxation between central galaxies and their host haloes is weaker for early galaxies or in older haloes, a trend opposite to expectation. The differences in $\alphac$ of early and late galaxies, however, are below 1.7$\sigma$ in all cases. Therefore, if the trend hints assembly bias effect, we do not have a statistically significant detection. The puzzle lies in the magnitude of the $\alphac$ constraints, which appear to be high compared to previous inference based on luminosity-threshold samples \citep[][]{Guo15c}. While $\alphac$ constraints for late galaxies largely fall into the range found in previous work, those for early galaxies are high, roughly in the range of $\alphac\sim$ 0.5--1.0. We suspect that this reflects assembly bias effect. For example, instead of the central galaxy motion in the frame of host haloes, the velocity bias constrained here captures the high pairwise velocity dispersion among early-formed haloes \citep[][]{Xu18}. In addition, the negative quadrupole on small scales measured in the EXT late galaxy samples may also signal assembly bias effect that cannot be described by a simple model.

To further investigate the possible assembly bias signatures hinted in our results, a halo model with assembly bias description \citep[e.g.][]{Hadzhiyska21,Yuan21} is necessary, and to reach a conclusive result a variety of clustering measurements may need to be modelled, which can include the redshift-space clustering, the lensing measurements, the environment-dependent clustering, the count-in-cell statistics, and void-related statistics \citep[e.g.][]{Wang19,Hadzhiyska21}.  In addition, such investigations will benefit from substantially larger galaxy samples than studied here, such as the Bright Galaxy Sample in the Dark Energy Spectroscopic Instrument survey \citep{DESI16}.

\section*{Acknowledgements}
KSM acknowledges the support by a fellowship from the Willard L. and Ruth P. Eccles Foundation. The work is supported by a seed grant at the University of Utah. The support and resources from the Center for High Performance Computing at the University of Utah are gratefully acknowledged. ZZ and HG acknowledge the support of the National Science Foundation of China (no. 11828302).

The CosmoSim database used in this paper is a service by the Leibniz-Institute for Astrophysics Potsdam (AIP). The MultiDark database was developed in cooperation with the Spanish MultiDark Consolider Project CSD2009-00064. The authors gratefully acknowledge the Gauss Centre for Supercomputing e.V. (\url{www.gauss-centre.eu}) and the Partnership for Advanced Supercomputing in Europe (PRACE, \url{www.prace-ri.eu}) for funding the MultiDark simulation project by providing computing time on the GCS Supercomputer SuperMUC at Leibniz Supercomputing Centre (LRZ, \url{www.lrz.de}).

\section*{Data Availability}
No new data were generated or analysed in support of this research.

\bibliographystyle{mnras}
\bibliography{ObsAB}

\begin{thebibliography}{}
\makeatletter
\relax
\def\mn@urlcharsother{\let\do\@makeother \do\$\do\&\do\#\do\^\do\_\do\%\do\~}
\def\mn@doi{\begingroup\mn@urlcharsother \@ifnextchar [ {\mn@doi@}
  {\mn@doi@[]}}
\def\mn@doi@[#1]#2{\def\@tempa{#1}\ifx\@tempa\@empty \href
  {http://dx.doi.org/#2} {doi:#2}\else \href {http://dx.doi.org/#2} {#1}\fi
  \endgroup}
\def\mn@eprint#1#2{\mn@eprint@#1:#2::\@nil}
\def\mn@eprint@arXiv#1{\href {http://arxiv.org/abs/#1} {{\tt arXiv:#1}}}
\def\mn@eprint@dblp#1{\href {http://dblp.uni-trier.de/rec/bibtex/#1.xml}
  {dblp:#1}}
\def\mn@eprint@#1:#2:#3:#4\@nil{\def\@tempa {#1}\def\@tempb {#2}\def\@tempc
  {#3}\ifx \@tempc \@empty \let \@tempc \@tempb \let \@tempb \@tempa \fi \ifx
  \@tempb \@empty \def\@tempb {arXiv}\fi \@ifundefined
  {mn@eprint@\@tempb}{\@tempb:\@tempc}{\expandafter \expandafter \csname
  mn@eprint@\@tempb\endcsname \expandafter{\@tempc}}}

\bibitem[\protect\citeauthoryear{{Abazajian} et~al.,}{{Abazajian}
  et~al.}{2005}]{Abazajian05}
{Abazajian} K.,  et~al., 2005, \mn@doi [\apj] {10.1086/429685}, \href
  {http://adsabs.harvard.edu/abs/2005ApJ...625..613A} {625, 613}

\bibitem[\protect\citeauthoryear{{Abazajian} et~al.,}{{Abazajian}
  et~al.}{2009}]{Abazajian09}
{Abazajian} K.~N.,  et~al., 2009, \mn@doi [\apjs]
  {10.1088/0067-0049/182/2/543}, \href
  {http://adsabs.harvard.edu/abs/2009ApJS..182..543A} {182, 543}

\bibitem[\protect\citeauthoryear{{Alongi}, {Bertelli}, {Bressan}, {Chiosi},
  {Fagotto}, {Greggio}  \& {Nasi}}{{Alongi} et~al.}{1993}]{Alongi93}
{Alongi} M.,  {Bertelli} G.,  {Bressan} A.,  {Chiosi} C.,  {Fagotto} F.,
  {Greggio} L.,   {Nasi} E.,  1993, \aaps, \href
  {http://adsabs.harvard.edu/abs/1993A%26AS...97..851A} {97, 851}

\bibitem[\protect\citeauthoryear{{Behroozi}, {Wechsler}  \& {Wu}}{{Behroozi}
  et~al.}{2013}]{Behroozi13a}
{Behroozi} P.~S.,  {Wechsler} R.~H.,   {Wu} H.-Y.,  2013, \mn@doi [\apj]
  {10.1088/0004-637X/762/2/109}, \href
  {http://adsabs.harvard.edu/abs/2013ApJ...762..109B} {762, 109}

\bibitem[\protect\citeauthoryear{{Bell}, {McIntosh}, {Katz}  \&
  {Weinberg}}{{Bell} et~al.}{2003}]{Bell03}
{Bell} E.~F.,  {McIntosh} D.~H.,  {Katz} N.,   {Weinberg} M.~D.,  2003, \mn@doi
  [\apjs] {10.1086/378847}, \href
  {http://adsabs.harvard.edu/abs/2003ApJS..149..289B} {149, 289}

\bibitem[\protect\citeauthoryear{{Berlind} et~al.,}{{Berlind}
  et~al.}{2003}]{Berlind03}
{Berlind} A.~A.,  et~al., 2003, \mn@doi [\apj] {10.1086/376517}, \href
  {http://adsabs.harvard.edu/abs/2003ApJ...593....1B} {593, 1}

\bibitem[\protect\citeauthoryear{{Blanton} et~al.,}{{Blanton}
  et~al.}{2005}]{Blanton05}
{Blanton} M.~R.,  et~al., 2005, \mn@doi [\aj] {10.1086/429803}, \href
  {http://adsabs.harvard.edu/abs/2005AJ....129.2562B} {129, 2562}

\bibitem[\protect\citeauthoryear{{Bond}, {Cole}, {Efstathiou}  \&
  {Kaiser}}{{Bond} et~al.}{1991}]{Bond91}
{Bond} J.~R.,  {Cole} S.,  {Efstathiou} G.,   {Kaiser} N.,  1991, \mn@doi
  [\apj] {10.1086/170520}, \href
  {http://adsabs.harvard.edu/abs/1991ApJ...379..440B} {379, 440}

\bibitem[\protect\citeauthoryear{{Bressan}, {Fagotto}, {Bertelli}  \&
  {Chiosi}}{{Bressan} et~al.}{1993}]{Bressan93}
{Bressan} A.,  {Fagotto} F.,  {Bertelli} G.,   {Chiosi} C.,  1993, \aaps, \href
  {http://adsabs.harvard.edu/abs/1993A%26AS..100..647B} {100, 647}

\bibitem[\protect\citeauthoryear{{Brinchmann}, {Charlot}, {White}, {Tremonti},
  {Kauffmann}, {Heckman}  \& {Brinkmann}}{{Brinchmann}
  et~al.}{2004}]{Brinchmann04}
{Brinchmann} J.,  {Charlot} S.,  {White} S.~D.~M.,  {Tremonti} C.,  {Kauffmann}
  G.,  {Heckman} T.,   {Brinkmann} J.,  2004, \mn@doi [\mnras]
  {10.1111/j.1365-2966.2004.07881.x}, \href
  {https://ui.adsabs.harvard.edu/abs/2004MNRAS.351.1151B} {351, 1151}

\bibitem[\protect\citeauthoryear{{Bruzual} \& {Charlot}}{{Bruzual} \&
  {Charlot}}{2003}]{Bruzual03}
{Bruzual} G.,  {Charlot} S.,  2003, \mn@doi [\mnras]
  {10.1046/j.1365-8711.2003.06897.x}, \href
  {http://adsabs.harvard.edu/abs/2003MNRAS.344.1000B} {344, 1000}

\bibitem[\protect\citeauthoryear{{Chabrier}}{{Chabrier}}{2003}]{Chabrier03}
{Chabrier} G.,  2003, \mn@doi [\pasp] {10.1086/376392}, \href
  {http://adsabs.harvard.edu/abs/2003PASP..115..763C} {115, 763}

\bibitem[\protect\citeauthoryear{{DESI Collaboration}}{{DESI
  Collaboration}}{2016}]{DESI16}
{DESI Collaboration} 2016, arXiv e-prints, \href
  {https://ui.adsabs.harvard.edu/abs/2016arXiv161100036D} {p. arXiv:1611.00036}

\bibitem[\protect\citeauthoryear{{Dalal}, {White}, {Bond}  \&
  {Shirokov}}{{Dalal} et~al.}{2008}]{Dalal08}
{Dalal} N.,  {White} M.,  {Bond} J.~R.,   {Shirokov} A.,  2008, \mn@doi [\apj]
  {10.1086/591512}, \href
  {https://ui.adsabs.harvard.edu/abs/2008ApJ...687...12D} {687, 12}

\bibitem[\protect\citeauthoryear{{Fagotto}, {Bressan}, {Bertelli}  \&
  {Chiosi}}{{Fagotto} et~al.}{1994a}]{Fagotto94a}
{Fagotto} F.,  {Bressan} A.,  {Bertelli} G.,   {Chiosi} C.,  1994a, \aaps,
  \href {http://adsabs.harvard.edu/abs/1994A%26AS..104..365F} {104, 365}

\bibitem[\protect\citeauthoryear{{Fagotto}, {Bressan}, {Bertelli}  \&
  {Chiosi}}{{Fagotto} et~al.}{1994b}]{Fagotto94b}
{Fagotto} F.,  {Bressan} A.,  {Bertelli} G.,   {Chiosi} C.,  1994b, \aaps,
  \href {http://adsabs.harvard.edu/abs/1994A%26AS..105...29F} {105, 29}

\bibitem[\protect\citeauthoryear{{Gao}, {Springel}  \& {White}}{{Gao}
  et~al.}{2005}]{Gao05}
{Gao} L.,  {Springel} V.,   {White} S.~D.~M.,  2005, \mn@doi [\mnras]
  {10.1111/j.1745-3933.2005.00084.x}, \href
  {http://adsabs.harvard.edu/abs/2005MNRAS.363L..66G} {363, L66}

\bibitem[\protect\citeauthoryear{{Girardi}, {Bressan}, {Chiosi}, {Bertelli}  \&
  {Nasi}}{{Girardi} et~al.}{1996}]{Girardi96}
{Girardi} L.,  {Bressan} A.,  {Chiosi} C.,  {Bertelli} G.,   {Nasi} E.,  1996,
  \aaps, \href {http://adsabs.harvard.edu/abs/1996A%26AS..117..113G} {117, 113}

\bibitem[\protect\citeauthoryear{{Guo}, {Zehavi}  \& {Zheng}}{{Guo}
  et~al.}{2012}]{Guo12}
{Guo} H.,  {Zehavi} I.,   {Zheng} Z.,  2012, \mn@doi [\apj]
  {10.1088/0004-637X/756/2/127}, \href
  {https://ui.adsabs.harvard.edu/abs/2012ApJ...756..127G} {756, 127}

\bibitem[\protect\citeauthoryear{{Guo} et~al.,}{{Guo} et~al.}{2014}]{Guo14}
{Guo} H.,  et~al., 2014, \mn@doi [\mnras] {10.1093/mnras/stu763}, \href
  {http://adsabs.harvard.edu/abs/2014MNRAS.441.2398G} {441, 2398}

\bibitem[\protect\citeauthoryear{{Guo} et~al.,}{{Guo} et~al.}{2015a}]{Guo15a}
{Guo} H.,  et~al., 2015a, \mn@doi [\mnras] {10.1093/mnras/stu2120}, \href
  {http://adsabs.harvard.edu/abs/2015MNRAS.446..578G} {446, 578}

\bibitem[\protect\citeauthoryear{{Guo} et~al.,}{{Guo} et~al.}{2015b}]{Guo15b}
{Guo} H.,  et~al., 2015b, \mn@doi [\mnras] {10.1093/mnras/stv1966}, \href
  {http://adsabs.harvard.edu/abs/2015MNRAS.453.4368G} {453, 4368}

\bibitem[\protect\citeauthoryear{{Guo} et~al.,}{{Guo} et~al.}{2015c}]{Guo15c}
{Guo} H.,  et~al., 2015c, \mn@doi [\mnras] {10.1093/mnras/stv1966}, \href
  {https://ui.adsabs.harvard.edu/abs/2015MNRAS.453.4368G} {453, 4368}

\bibitem[\protect\citeauthoryear{{Hadzhiyska}, {Bose}, {Eisenstein}  \&
  {Hernquist}}{{Hadzhiyska} et~al.}{2021}]{Hadzhiyska21}
{Hadzhiyska} B.,  {Bose} S.,  {Eisenstein} D.,   {Hernquist} L.,  2021, \mn@doi
  [\mnras] {10.1093/mnras/staa3776}, \href
  {https://ui.adsabs.harvard.edu/abs/2021MNRAS.501.1603H} {501, 1603}

\bibitem[\protect\citeauthoryear{{Han}, {Li}, {Jing}, {Nishimichi}, {Wang}  \&
  {Jiang}}{{Han} et~al.}{2019}]{Han19}
{Han} J.,  {Li} Y.,  {Jing} Y.,  {Nishimichi} T.,  {Wang} W.,   {Jiang} C.,
  2019, \mn@doi [\mnras] {10.1093/mnras/sty2822}, \href
  {https://ui.adsabs.harvard.edu/abs/2019MNRAS.482.1900H} {482, 1900}

\bibitem[\protect\citeauthoryear{{Hartlap}, {Simon}  \& {Schneider}}{{Hartlap}
  et~al.}{2007}]{Hartlap07}
{Hartlap} J.,  {Simon} P.,   {Schneider} P.,  2007, \mn@doi [\aap]
  {10.1051/0004-6361:20066170}, \href
  {https://ui.adsabs.harvard.edu/abs/2007A&A...464..399H} {464, 399}

\bibitem[\protect\citeauthoryear{{Hearin} \& {Watson}}{{Hearin} \&
  {Watson}}{2013}]{HW13}
{Hearin} A.~P.,  {Watson} D.~F.,  2013, \mn@doi [\mnras]
  {10.1093/mnras/stt1374}, \href
  {https://ui.adsabs.harvard.edu/abs/2013MNRAS.435.1313H} {435, 1313}

\bibitem[\protect\citeauthoryear{{Klypin}, {Yepes}, {Gottl{\"o}ber}, {Prada}
  \& {He{\ss}}}{{Klypin} et~al.}{2016}]{Klypin16}
{Klypin} A.,  {Yepes} G.,  {Gottl{\"o}ber} S.,  {Prada} F.,   {He{\ss}} S.,
  2016, \mn@doi [\mnras] {10.1093/mnras/stw248}, \href
  {https://ui.adsabs.harvard.edu/abs/2016MNRAS.457.4340K} {457, 4340}

\bibitem[\protect\citeauthoryear{{Landy} \& {Szalay}}{{Landy} \&
  {Szalay}}{1993}]{Landy93}
{Landy} S.~D.,  {Szalay} A.~S.,  1993, \mn@doi [\apj] {10.1086/172900}, \href
  {https://ui.adsabs.harvard.edu/abs/1993ApJ...412...64L} {412, 64}

\bibitem[\protect\citeauthoryear{{Lin}, {Mandelbaum}, {Huang}, {Huang},
  {Dalal}, {Diemer}, {Jian}  \& {Kravtsov}}{{Lin} et~al.}{2016}]{Lin16}
{Lin} Y.-T.,  {Mandelbaum} R.,  {Huang} Y.-H.,  {Huang} H.-J.,  {Dalal} N.,
  {Diemer} B.,  {Jian} H.-Y.,   {Kravtsov} A.,  2016, \mn@doi [\apj]
  {10.3847/0004-637X/819/2/119}, \href
  {http://adsabs.harvard.edu/abs/2016ApJ...819..119L} {819, 119}

\bibitem[\protect\citeauthoryear{{Luo} et~al.,}{{Luo} et~al.}{2017}]{Wentao17}
{Luo} W.,  et~al., 2017, \mn@doi [\apj] {10.3847/1538-4357/836/1/38}, \href
  {https://ui.adsabs.harvard.edu/abs/2017ApJ...836...38L} {836, 38}

\bibitem[\protect\citeauthoryear{{McCarthy}, {Zheng}  \& {Guo}}{{McCarthy}
  et~al.}{2019}]{McCarthy2019}
{McCarthy} K.~S.,  {Zheng} Z.,   {Guo} H.,  2019, \mn@doi [\mnras]
  {10.1093/mnras/stz1461}, \href
  {https://ui.adsabs.harvard.edu/abs/2019MNRAS.487.2424M} {487, 2424}

\bibitem[\protect\citeauthoryear{{Miyatake}, {More}, {Takada}, {Spergel},
  {Mandelbaum}, {Rykoff}  \& {Rozo}}{{Miyatake} et~al.}{2016}]{Miyatake16}
{Miyatake} H.,  {More} S.,  {Takada} M.,  {Spergel} D.~N.,  {Mandelbaum} R.,
  {Rykoff} E.~S.,   {Rozo} E.,  2016, \mn@doi [Physical Review Letters]
  {10.1103/PhysRevLett.116.041301}, \href
  {http://adsabs.harvard.edu/abs/2016PhRvL.116d1301M} {116, 041301}

\bibitem[\protect\citeauthoryear{{More}, {van den Bosch}, {Cacciato}, {Skibba},
  {Mo}  \& {Yang}}{{More} et~al.}{2011}]{More11}
{More} S.,  {van den Bosch} F.~C.,  {Cacciato} M.,  {Skibba} R.,  {Mo} H.~J.,
  {Yang} X.,  2011, \mn@doi [\mnras] {10.1111/j.1365-2966.2010.17436.x}, \href
  {http://adsabs.harvard.edu/abs/2011MNRAS.410..210M} {410, 210}

\bibitem[\protect\citeauthoryear{{Paranjape}, {Hahn}  \& {Sheth}}{{Paranjape}
  et~al.}{2018}]{Paranjape18}
{Paranjape} A.,  {Hahn} O.,   {Sheth} R.~K.,  2018, \mn@doi [\mnras]
  {10.1093/mnras/sty496}, \href
  {https://ui.adsabs.harvard.edu/abs/2018MNRAS.476.3631P} {476, 3631}

\bibitem[\protect\citeauthoryear{{Reid}, {Seo}, {Leauthaud}, {Tinker}  \&
  {White}}{{Reid} et~al.}{2014}]{Reid14}
{Reid} B.~A.,  {Seo} H.-J.,  {Leauthaud} A.,  {Tinker} J.~L.,   {White} M.,
  2014, \mn@doi [\mnras] {10.1093/mnras/stu1391}, \href
  {https://ui.adsabs.harvard.edu/abs/2014MNRAS.444..476R} {444, 476}

\bibitem[\protect\citeauthoryear{{Salcedo} et~al.,}{{Salcedo}
  et~al.}{2020}]{Salcedo20}
{Salcedo} A.~N.,  et~al., 2020, arXiv e-prints, \href
  {https://ui.adsabs.harvard.edu/abs/2020arXiv201004176S} {p. arXiv:2010.04176}

\bibitem[\protect\citeauthoryear{{Sheth} \& {Tormen}}{{Sheth} \&
  {Tormen}}{2004}]{Sheth04}
{Sheth} R.~K.,  {Tormen} G.,  2004, \mn@doi [\mnras]
  {10.1111/j.1365-2966.2004.07733.x}, \href
  {http://adsabs.harvard.edu/abs/2004MNRAS.350.1385S} {350, 1385}

\bibitem[\protect\citeauthoryear{{Tojeiro}, {Heavens}, {Jimenez}  \&
  {Panter}}{{Tojeiro} et~al.}{2007}]{Tojeiro07}
{Tojeiro} R.,  {Heavens} A.~F.,  {Jimenez} R.,   {Panter} B.,  2007, \mn@doi
  [\mnras] {10.1111/j.1365-2966.2007.12323.x}, \href
  {http://adsabs.harvard.edu/abs/2007MNRAS.381.1252T} {381, 1252}

\bibitem[\protect\citeauthoryear{{Tojeiro}, {Wilkins}, {Heavens}, {Panter}  \&
  {Jimenez}}{{Tojeiro} et~al.}{2009}]{Tojeiro09}
{Tojeiro} R.,  {Wilkins} S.,  {Heavens} A.~F.,  {Panter} B.,   {Jimenez} R.,
  2009, \mn@doi [\apjs] {10.1088/0067-0049/185/1/1}, \href
  {http://adsabs.harvard.edu/abs/2009ApJS..185....1T} {185, 1}

\bibitem[\protect\citeauthoryear{{Vakili} \& {Hahn}}{{Vakili} \&
  {Hahn}}{2019}]{Vakili19}
{Vakili} M.,  {Hahn} C.,  2019, \mn@doi [\apj] {10.3847/1538-4357/aaf1a1},
  \href {https://ui.adsabs.harvard.edu/abs/2019ApJ...872..115V} {872, 115}

\bibitem[\protect\citeauthoryear{{Wang} et~al.,}{{Wang} et~al.}{2019}]{Wang19}
{Wang} K.,  et~al., 2019, \mn@doi [\mnras] {10.1093/mnras/stz1733}, \href
  {https://ui.adsabs.harvard.edu/abs/2019MNRAS.488.3541W} {488, 3541}

\bibitem[\protect\citeauthoryear{{Wechsler} \& {Tinker}}{{Wechsler} \&
  {Tinker}}{2018}]{Wechsler18}
{Wechsler} R.~H.,  {Tinker} J.~L.,  2018, \mn@doi [\araa]
  {10.1146/annurev-astro-081817-051756}, \href
  {https://ui.adsabs.harvard.edu/abs/2018ARA&A..56..435W} {56, 435}

\bibitem[\protect\citeauthoryear{{Xu} \& {Zheng}}{{Xu} \& {Zheng}}{2018}]{Xu18}
{Xu} X.,  {Zheng} Z.,  2018, \mn@doi [\mnras] {10.1093/mnras/sty1547}, \href
  {https://ui.adsabs.harvard.edu/abs/2018MNRAS.479.1579X} {479, 1579}

\bibitem[\protect\citeauthoryear{{Xu} \& {Zheng}}{{Xu} \& {Zheng}}{2020}]{Xu20}
{Xu} X.,  {Zheng} Z.,  2020, \mn@doi [\mnras] {10.1093/mnras/staa009}, \href
  {https://ui.adsabs.harvard.edu/abs/2020MNRAS.492.2739X} {492, 2739}

\bibitem[\protect\citeauthoryear{{Xu}, {Zheng}, {Guo}, {Zu}, {Zehavi}  \&
  {Weinberg}}{{Xu} et~al.}{2018}]{XuHJ18}
{Xu} H.,  {Zheng} Z.,  {Guo} H.,  {Zu} Y.,  {Zehavi} I.,   {Weinberg} D.~H.,
  2018, \mn@doi [\mnras] {10.1093/mnras/sty2615}, \href
  {https://ui.adsabs.harvard.edu/abs/2018MNRAS.481.5470X} {481, 5470}

\bibitem[\protect\citeauthoryear{{Yang}, {Mo}  \& {van den Bosch}}{{Yang}
  et~al.}{2003}]{Yang03}
{Yang} X.,  {Mo} H.~J.,   {van den Bosch} F.~C.,  2003, \mn@doi [\mnras]
  {10.1046/j.1365-8711.2003.06254.x}, \href
  {http://adsabs.harvard.edu/abs/2003MNRAS.339.1057Y} {339, 1057}

\bibitem[\protect\citeauthoryear{{Yang}, {Mo}, {van den Bosch}, {Pasquali},
  {Li}  \& {Barden}}{{Yang} et~al.}{2007}]{Yang07}
{Yang} X.,  {Mo} H.~J.,  {van den Bosch} F.~C.,  {Pasquali} A.,  {Li} C.,
  {Barden} M.,  2007, \mn@doi [\apj] {10.1086/522027}, \href
  {http://adsabs.harvard.edu/abs/2007ApJ...671..153Y} {671, 153}

\bibitem[\protect\citeauthoryear{{Yang}, {Mo}  \& {van den Bosch}}{{Yang}
  et~al.}{2008}]{Yang08}
{Yang} X.,  {Mo} H.~J.,   {van den Bosch} F.~C.,  2008, \mn@doi [\apj]
  {10.1086/528954}, \href {http://adsabs.harvard.edu/abs/2008ApJ...676..248Y}
  {676, 248}

\bibitem[\protect\citeauthoryear{{Ye}, {Guo}, {Zheng}  \& {Zehavi}}{{Ye}
  et~al.}{2017}]{Ye17}
{Ye} J.-N.,  {Guo} H.,  {Zheng} Z.,   {Zehavi} I.,  2017, \mn@doi [\apj]
  {10.3847/1538-4357/aa70e7}, \href
  {https://ui.adsabs.harvard.edu/abs/2017ApJ...841...45Y} {841, 45}

\bibitem[\protect\citeauthoryear{{York} et~al.,}{{York} et~al.}{2000}]{York00}
{York} D.~G.,  et~al., 2000, \mn@doi [\aj] {10.1086/301513}, \href
  {http://adsabs.harvard.edu/abs/2000AJ....120.1579Y} {120, 1579}

\bibitem[\protect\citeauthoryear{{Yuan}, {Hadzhiyska}, {Bose}, {Eisenstein}  \&
  {Guo}}{{Yuan} et~al.}{2021}]{Yuan21}
{Yuan} S.,  {Hadzhiyska} B.,  {Bose} S.,  {Eisenstein} D.~J.,   {Guo} H.,
  2021, \mn@doi [\mnras] {10.1093/mnras/stab235}, \href
  {https://ui.adsabs.harvard.edu/abs/2021MNRAS.502.3582Y} {502, 3582}

\bibitem[\protect\citeauthoryear{{Zehavi} et~al.,}{{Zehavi}
  et~al.}{2005}]{Zehavi05}
{Zehavi} I.,  et~al., 2005, \mn@doi [\apj] {10.1086/431891}, \href
  {http://adsabs.harvard.edu/abs/2005ApJ...630....1Z} {630, 1}

\bibitem[\protect\citeauthoryear{{Zehavi} et~al.,}{{Zehavi}
  et~al.}{2011}]{Zehavi11}
{Zehavi} I.,  et~al., 2011, \mn@doi [\apj] {10.1088/0004-637X/736/1/59}, \href
  {https://ui.adsabs.harvard.edu/abs/2011ApJ...736...59Z} {736, 59}

\bibitem[\protect\citeauthoryear{{Zentner}, {Hearin}  \& {van den
  Bosch}}{{Zentner} et~al.}{2014}]{Zentner14}
{Zentner} A.~R.,  {Hearin} A.~P.,   {van den Bosch} F.~C.,  2014, \mn@doi
  [\mnras] {10.1093/mnras/stu1383}, \href
  {http://adsabs.harvard.edu/abs/2014MNRAS.443.3044Z} {443, 3044}

\bibitem[\protect\citeauthoryear{{Zentner}, {Hearin}, {van den Bosch}, {Lange}
  \& {Villarreal}}{{Zentner} et~al.}{2019}]{Zentner19}
{Zentner} A.~R.,  {Hearin} A.,  {van den Bosch} F.~C.,  {Lange} J.~U.,
  {Villarreal} A.,  2019, \mn@doi [\mnras] {10.1093/mnras/stz470}, \href
  {https://ui.adsabs.harvard.edu/abs/2019MNRAS.485.1196Z} {485, 1196}

\bibitem[\protect\citeauthoryear{{Zhai} et~al.,}{{Zhai} et~al.}{2019}]{Zhai19}
{Zhai} Z.,  et~al., 2019, \mn@doi [\apj] {10.3847/1538-4357/ab0d7b}, \href
  {https://ui.adsabs.harvard.edu/abs/2019ApJ...874...95Z} {874, 95}

\bibitem[\protect\citeauthoryear{{Zheng}}{{Zheng}}{2004}]{Zheng04}
{Zheng} Z.,  2004, \mn@doi [\apj] {10.1086/421542}, \href
  {http://adsabs.harvard.edu/abs/2004ApJ...610...61Z} {610, 61}

\bibitem[\protect\citeauthoryear{{Zheng} \& {Guo}}{{Zheng} \&
  {Guo}}{2016}]{Zheng16}
{Zheng} Z.,  {Guo} H.,  2016, \mn@doi [\mnras] {10.1093/mnras/stw523}, \href
  {https://ui.adsabs.harvard.edu/abs/2016MNRAS.458.4015Z} {458, 4015}

\bibitem[\protect\citeauthoryear{{Zheng} \& {Weinberg}}{{Zheng} \&
  {Weinberg}}{2007}]{Zheng07}
{Zheng} Z.,  {Weinberg} D.~H.,  2007, \mn@doi [\apj] {10.1086/512151}, \href
  {http://adsabs.harvard.edu/abs/2007ApJ...659....1Z} {659, 1}

\bibitem[\protect\citeauthoryear{{Zheng} et~al.,}{{Zheng}
  et~al.}{2005}]{Zheng05}
{Zheng} Z.,  et~al., 2005, \mn@doi [\apj] {10.1086/466510}, \href
  {http://adsabs.harvard.edu/abs/2005ApJ...633..791Z} {633, 791}

\bibitem[\protect\citeauthoryear{{Zu}, {Mandelbaum}, {Simet}, {Rozo}  \&
  {Rykoff}}{{Zu} et~al.}{2017}]{Zu17}
{Zu} Y.,  {Mandelbaum} R.,  {Simet} M.,  {Rozo} E.,   {Rykoff} E.~S.,  2017,
  \mn@doi [\mnras] {10.1093/mnras/stx1264}, \href
  {http://adsabs.harvard.edu/abs/2017MNRAS.470..551Z} {470, 551}

\bibitem[\protect\citeauthoryear{{Zu} et~al.,}{{Zu} et~al.}{2021}]{Zu21}
{Zu} Y.,  et~al., 2021, \mn@doi [\mnras] {10.1093/mnras/stab1712}, \href
  {https://ui.adsabs.harvard.edu/abs/2021MNRAS.505.5117Z} {505, 5117}

\bibitem[\protect\citeauthoryear{{van den Bosch}, {Mo}  \& {Yang}}{{van den
  Bosch} et~al.}{2003}]{vandenBosch03}
{van den Bosch} F.~C.,  {Mo} H.~J.,   {Yang} X.,  2003, \mn@doi [\mnras]
  {10.1046/j.1365-8711.2003.07012.x}, \href
  {http://adsabs.harvard.edu/abs/2003MNRAS.345..923V} {345, 923}

\bibitem[\protect\citeauthoryear{{van den Bosch}, {More}, {Cacciato}, {Mo}  \&
  {Yang}}{{van den Bosch} et~al.}{2013}]{vandenBosch13}
{van den Bosch} F.~C.,  {More} S.,  {Cacciato} M.,  {Mo} H.,   {Yang} X.,
  2013, \mn@doi [\mnras] {10.1093/mnras/sts006}, \href
  {http://adsabs.harvard.edu/abs/2013MNRAS.430..725V} {430, 725}

\makeatother
\end{thebibliography}

\bsp    % typesetting comment
\label{lastpage}
\end{document}